\documentstyle[preprint,prb,aps,amsmath,amssymb,amsthm,epsf]{revtex}

\numberwithin{equation}{section}

\newtheorem{lem}{Lemma}[section]
\newtheorem{theo}{Theorem}[section]

\newtheorem{defi}{Definition}[section]

\DeclareMathOperator*{\slim}{s--lim}

\DeclareMathOperator{\Ret}{Re}
\DeclareMathOperator{\Imt}{Im}
\renewcommand{\Im}{\Imt}

\DeclareMathOperator{\sgn}{sgn}
\DeclareMathOperator{\supp}{supp}

\newcommand{\HH}{{\mathbb{H}}}
\newcommand{\R}{{\mathbb{R}}}
\newcommand{\C}{{\mathbb{C}}}
\newcommand{\Z}{{\mathbb{Z}}}

\newcommand{\cL}{{\cal L}}

\newcommand{\cC}{{\cal C}}

\newcommand{\cF}{{\cal F}}

\newcommand{\id}{{\mathbf 1}}

\newcommand{\cone}{\C^\star / \Z_2}             

\newcommand{\nC}{{^n{\mathbb C}}}               
\newcommand{\NC}{{^n\tilde{\mathbb C}}}         

\newcommand{\ket}[1]{\left.\left\lvert #1 \right.\right\rangle}
\newcommand{\skal}[2]{\left\langle #1\mid #2 \right\rangle}
\newcommand{\matrele}[3]{\left\langle#1\,\left\lvert\,#2\,\right\rvert\,#3\right\rangle}

\newcommand{\norm}[1]{{\left\lVert#1\right\rVert}}  
\newcommand{\modulus}[1]{{\,\left\lvert#1\right\rvert\,}}

\newcommand{\Absatz}{\vspace{5mm} \par \noindent}          

\title{Two-particle scattering theory for anyons}

\author{C.~Korff$^{1,}$\thanks{christian.korff@physik.fu-berlin.de},  
G.~Lang$^{2,}$\thanks{lang@math.tu-berlin.de}, 
R.~Schrader$^{1,}$\thanks{robert.schrader@physik.fu-berlin.de}}
\address{$^1$Institut f\"ur Theoretische Physik, Freie Universit\"at Berlin\\
 Arnimallee 14, 14195 Berlin, Germany}
\address{
$^2$Fachbereich Mathematik, Technische Universit\"at Berlin\\
Strasse des 17. Juni 136, 10623 Berlin,Germany}

\tighten
\tightenlines

\begin{document}

\newdimen\randvollbild
\newdimen\abstandbildseite
\newdimen\breiteseitebild
\newdimen\bildbreite

\randvollbild1cm
\abstandbildseite0.5cm
\breiteseitebild9.5cm
\bildbreite \textwidth
\advance \bildbreite by -\randvollbild
\advance \bildbreite by -\randvollbild



\newcounter{temp}

\makeatletter
\def\pagereftocounter#1#2{%
 \setcounter{#2}{\expandafter\@setref\csname r@#1\endcsname\@secondoftwo{#1} }}
\makeatother

\newcommand{\bildeinbinden}[3]{
\begin{figure}[tbp]
  \vspace{0.3cm}
  \centerline{\epsffile{#1}}
  \renewcommand{\normalsize}{\footnotesize \bf}
  \centerline{
    \begin{minipage}{\bildbreite}\caption[#2]{\label{#2}{\rm #3}}
    \end{minipage}
    }
  \renewcommand{\normalsize}{\standard}
\end{figure}}


\newcommand{\bildeinbindencm}[4]{
\begin{figure}[htb]
  \vspace{0.3cm}
  \centerline{
    \epsfxsize=#1
    \epsffile{#2}    }
  \renewcommand{\normalsize}{\footnotesize \bf}
  \centerline{
    \begin{minipage}{\bildbreite}\caption[#3]{\label{#3}{\rm #4}}
    \end{minipage}
    }
  \renewcommand{\normalsize}{\standard}
\end{figure}}

\maketitle


\begin{abstract}
  We consider potential scattering theory of a nonrelativistic quantum
  mechanical 2-particle system 
  in $\R^{2}$ with anyon statistics. Sufficient conditions are given which
  guarantee the existence of M{\o}ller operators and the unitarity of
  the $S$--matrix. As examples the rotationally invariant potential
  well and the $\delta$--function potential are discussed in detail. In
  case of a general rotationally invariant potential the angular
  momentum decomposition leads to a theory of Jost functions.  The
  anyon statistics parameter gives rise to an interpolation for
  angular momenta analogous to the Regge trajectories for complex
  angular momenta. Levinson's theorem is adapted to the present
  context. In particular we find that in case of a zero energy
  resonance the statistics parameter can be determined from the
  scattering phase.
\end{abstract}

\draft




\section{Introduction}
In recent years the theory of identical quantum mechanical particles
with braid group statistics has received increasing attention (for a
recent review and references on anyon physics see e.g. \onlinecite{Lerda92,Montonen94}). The
original observation that in two dimensional configuration space
$\R^2$ identical particles may obey statistics different from Bose or
Fermi statistics is due to J.M. Leinaas and J. Myrheim \cite{Leinaas77}.
They based the notion of quantum statistics on the topological structure of 
the classical configuration space for identical particles. The relevant symmetry group is then shown 
to be the braid group which replaces the permutation group. Models for
particles with a one--dimensional representation of the braid group
were first discussed by F.~Wilczek, who coined the name anyons for
particles with these new  statistics \cite{Wilczek82} (see also
\onlinecite{Goldin81,Goldin83,Goldin85}). Wilczek suggested the following physical 
picture of anyons: Magnetic flux tubes (vortices) are attached to
either charged  bosons or fermions. The latter then give rise to
arbitrary Aharonov--Bohm phases when transported along paths
exchanging the particle positions. The magnetic flux tubes are
described by long range gauge potentials whose curvature vanishes and 
can be related to Chern--Simons theory
\onlinecite{Wilczek,Wilczek82}. This point of view was taken 
up by several authors (e.g. \onlinecite{Arovas85,Hansson88}) leading to a second
quantized version of anyons obtained by coupling a
Chern--Simons U(1)--gauge potential to a matter field \cite{Jackiw90}.\\ 
Also the general case, where the finite dimensional representation of the 
braid group is not one--dimensional, has been considered. The corresponding 
particles  are called ``plektons'' \cite{Fredenhagen92,Fredenhagen89}. 
The relevance of braid group statistics in conformal
quantum field theory was realized by Tsuchiya and Kanie
\cite{Tsuchiya88} and in algebraic quantum field theory by
J.~Fr\"ohlich \cite{Froehlich89} and K.~Fredenhagen, K.--H.~Rehren and
B.~Schroer \cite{Fredenhagen89,Fredenhagen92}.\\
However, in the discussion of particles with braid group statistics the main focus is 
on anyons (abelian statistics), since in many particle theory they might provide an explanation 
for the fractional quantum Hall effect and for high $T_c$ superconductivity (
for a review see e.g. \onlinecite{Girvin87,Stone92} and \onlinecite{Wilczek90,Lykken91}). 
In this article we want do discuss non--relativistic two--particle
potential scattering theory for anyons. This is done in the framework
of ordinary quantum mechanics, i.e. we insert a gauge potential of the
Aharonov--Bohm type in the center of mass 
Hamiltonian. Emphasis is then put on showing
how well known techniques of scattering theory extend to the case
of anyon statistics. The motivation is twofold. On the one hand
scattering theory is a powerful tool in spectral analysis and thus
might be helpful for a better understanding of fractional
statistics. On the other hand scattering data can be used to compute
the virial coefficients of an interacting anyon gas
\cite{Beth37}. Hence, it is possible to infer bulk properties from
scattering theory. The latter are of main interest in the
investigation of the above mentioned phenomena. \\ 
The first calculation of the second virial coefficient for the 
two--particle anyon system was done by Arovas et
al. \cite{Arovas85}. They considered the case when only the statistics
is present corresponding to the case of
Aharonov--Bohm scattering \cite{Aharonov59,Ruijsenaars83}. See
\onlinecite{Kim98} and references therein for recent articles on the second virial
coefficient of {\em interacting} anyon systems. In a forthcoming publication 
we intend to apply the results of this article to the calculation of
the second virial coefficient.
\Absatz 
The article is organized as follows. In
section \ref{sec:qm} we review the quantum mechanics of two ``free''
anyons in order to introduce our notation and to keep the paper
self--contained. In particular, we give the energy eigenfunctions, the
resolvent (Green's function), and the propagator. We also 
recapitulate the relation of the free anyon system and Aharonov--Bohm 
scattering which will be used in our discussion of the differential
cross--section in section \ref{sec:cross}. In appendix \ref{sec:bun} we
recall the equivalent differential geometric formulation in terms of
vector bundles, which in the case of anyons are line bundles. In particular, 
we recall that there exists a canonical hermitian 
connection encoding the statistics such that the ``free'' Hamiltonian is 
the canonically associated Bochner Laplacean. \\
In section \ref{sec:pot} we consider scattering theory for the
interacting anyon system obtained by adding a potential to the center
of mass Hamiltonian. We give
sufficient conditions for the existence of the M\o ller operators,
which also cover the non spherical symmetric
case. Applying the Kuroda--Birman theorem \cite{Kuroda62} we derive
the unitarity of the resulting $S$--matrix.\\
In section \ref{sec:cross} we discuss the differential cross--section
with the modifications necessary to accommodate anyon statistics. 
We show that the scattering amplitude splits 
into two parts, one describing the effects of the statistics the other
the interaction represented by the potential.\\
In section \ref{sec:Jost} Jost functions are introduced which depend
on the statistics parameter for anyons. The latter enters in the form
of continuous angular momentum. This establishes a connection with
Regge trajectories in the theory of complex angular momenta. We conclude
section \ref{sec:Jost} by showing how Levinson's theorem, which
relates the scattering phase shift to the number of bound states,
carries over to the present situation. In case that the Jost
function vanishes at zero energy we derive an explicit formula giving
the statistics parameter in terms of the scattering phase shift.\\
Section \ref{sec:delta} and \ref{sec:step} are devoted to explicitly
solvable examples. In section \ref{sec:delta} 
we examine the $\delta$-potential. This case also figures under the
name of anyons without hard--core condition. The corresponding resolvent is calculated 
in closed form in appendix \ref{sec:res} and the bound state problem is then 
considered. We also remark on the modification of Levinson's theorem
and find an additional formula relating the statistics parameter to
the scattering phase. In
section \ref{sec:step} we discuss the square well potential. The Jost
function is calculated and Regge trajectories are plotted
which show the dependence of the point spectrum on angular momentum
and the statistics parameter. In section \ref{sec:delta} and
\ref{sec:step} we provide numerical examples for the differential
cross--section which display the interpolation between Bose and Fermi
statistics when the statistics parameter for anyons is varied.\Absatz
The results presented here are based in part on the diploma thesis of 
two of the authors (C.K.\cite{Korff97} and G.L.\cite{Lang97}).\Absatz  
Throughout the article we will work in atomic units, $\hbar=e=m=1$,
where $m$ denotes the mass of the particles. In particular this sets
the reduced mass of a two--particle system equal to $1/2$. In
estimates we make the convention that $C,C(\epsilon)$ etc. denote generic
constants depending on $\epsilon$ etc.

\section{Quantum Mechanics of Two Anyons}
\label{sec:qm}
The theory of identical particles with statistics differing from Bose
or Fermi statistics may show up when the configuration (or momentum)
space of one particle is the two dimensional Euclidean space.
Henceforth we will often use the complex plane $\C$ to describe such a
space. For the configuration space $\C\times\C$ of two non identical
particles with points labeled by $(z_1,z_2)$ the relative coordinate
$z=z_1-z_2$ changes into $-z$ if the coordinates $z_1$ and $z_2$ of
the two particles are interchanged. The basic observation of Leinaas
and Myrheim was, that by leaving out the case where the two particles
are at the same point, i.e. where $z=0$, the configuration space in
the center of mass frame of two identical particles should be the
space obtained from $\C^{\star}=\C\setminus\{0\}$ by identifying the
points $z$ and $-z$.  This space is therefore the orbit space $\cone$,
where $\Z_{2}=\{+1,-1\}$ acts in the obvious way as a transformation
group on $\C^{\star}$. This space is also obtained from the closed
upper half--plane $\HH$ in $\C$ minus the origin, i.e the set
$\HH\setminus\{0\}$, by identifying the points $x$ and $-x$ on the
real axis. Geometrically this leads to a cone with removed apex as
configuration space. The obvious choice of polar coordinates
$(r,\theta) \in \R^+ \times [0,\pi)$ on $\HH$ carries over to the cone
$\cone$. We take the state vectors of the system to be the square
integrable functions on the cone $\cL^2(\cone)$ or equivalently the
$\pi$--periodic functions on the punctured plane $\C^\star$. The fact
that the cone's apex or respectively the plane's origin is removed,
allows for the particles to carry flux--tubes. This corresponds to the
physical picture introduced by F.~Wilczek (see e.g.
\onlinecite{Wilczek82,Wilczek90}), who named such particles {\em
  anyons}. The flux--tubes are taken into account by
inserting a gauge potential of Aharonov--Bohm type 
\begin{equation}\label{qm0}
A_\alpha=\frac{\alpha}{r} \, e_\theta,\qquad\alpha\in[0,1]
\end{equation}
into the center of mass Hamiltonian. Here $\alpha$ is the so called statistics
parameter and $e_\theta$ denotes the unit vector corresponding
to the polar angle. Choosing units in such a way that
$\hbar=1$ and setting the mass of the particles to one, the resulting 
Hamiltonian has the form
\begin{displaymath}\label{qm1} 
  H_0(\alpha) = - (\nabla+iA_{\alpha})^2 =
-\frac{\partial^2}{\partial r^2}- \frac{1}{r}\frac{\partial}{\partial
  r} -\frac{1}{r^2} \left(\frac{\partial}{\partial \theta}+i\alpha\right)^2  
\end{displaymath} 
in polar coordinates. This operator will be considered to be the free
Hamiltonian in the center of mass frame for a two--particle anyon
system with statistics parameter $\alpha\in[0,1]$. 
If $\alpha=0$ the particles actually behave like bosons, while for
$\alpha=1$ they behave like fermions. In appendix~\ref{sec:bun} we give
a short review of a mathematically precise formulation of this model in
terms of vector bundles, following \onlinecite{Mund92,Mund95} (see also
\onlinecite{Horvathy89,Imbo90}. There we
also argue why $\alpha=0,1$ corresponds to bosons and fermions
respectively and why we may restrict the parameter $\alpha$ to the
interval $[0,1]$.
 
We now determine the spectrum and the eigenfunctions of 
$H_{0}(\alpha)$. We start with the decomposition 
\begin{displaymath}\label{qm2} 
  \cL^2(\cone) \cong\cL^2(\R^+,r\,dr) \otimes \cL^2(S^{1}, 
  d\theta)   \ ,  
\end{displaymath} 
where the points in $S^{1}$, the unit circle in $\C$, are 
parameterized as $\exp (2i\theta)$ with $0\le\theta <\pi$. This leads 
to the decomposition 
\begin{equation}\label{qm3} 
  \cL^2(\cone)\cong \underset{m\in\Z}{\oplus} {\frak h}_{2m} \ , 
  \qquad {\frak h}_{2m}=\cL^2(\R^{+},rdr)\otimes\left\{  
    \frac{e^{2im\theta}}{\sqrt{\pi}}\right\}. 
\end{equation} 
This decomposition is of course related to the following fact. The 
rotation group $SO(2)\cong U(1)$ maps $\C^{\star}$ into itself and 
commutes with the action of $\Z_{2}$, thus acts on $\cone$ and defines a unitary 
action on $L^2(\cone)$. Also $H_{0}(\alpha)$ commutes with this action of U(1), 
it is diagonal w.r.t. the decomposition, i.e. 
\begin{displaymath}\label{qm5} 
  H_{0}(\alpha )=\underset{m\in\Z}{\oplus} \,H_{0,2m}(\alpha ) 
\end{displaymath} 
with 
\begin{displaymath}\label{qm6} 
  H_{0,2m}(\alpha )\,=\, -\left \{ \frac{d^2}{dr^2}+\frac{1}{r}\, 
    \frac{d}{dr} -\frac{(2m+\alpha )^2}{r^2}\right\} 
\end{displaymath} 
on $\cL^2(\R^{+},r\,dr)\cong\,{\frak h}_{2m}$.  To find all solutions of the stationary 
Schr{\"o}dinger equation for $H_0(\alpha)$ it therefore suffices to find the solutions of the 
Bessel equation for each $m\in\Z$ 
\begin{equation}
  \label{qm7} 
  \left\{\frac{d^2}{dr}+ \frac{1}{r}\frac{d}{dr}- 
  \frac{(2m+\alpha)^2}{r^2} +E\right\} \,R(r)=0 \ . 
\end{equation} 
For definiteness we choose those solutions of equations~\eqref{qm7}
which are regular near $r=0$.  This gives the following improper
eigenfunctions
\begin{displaymath}
  \label{qm8} 
  \phi_{\alpha ;m,E}(r,\theta)=\frac{e^{2im\theta}}{\sqrt{2\pi}} \cdot 
  J_{\mid   2m+\alpha \mid}(\sqrt {E}r) \ .   
\end{displaymath} 
In what follows we will use the notations 
\begin{displaymath}
  \label{qm9} 
  k:=\sqrt{E},\qquad     \mu:=\mid2m+\alpha\mid . 
\end{displaymath} 
The identity  
\begin{displaymath}
  \label{qm10} 
  \int_{0}^{\infty}J_{\mu}(at)J_{\mu}(bt)\,tdt=\frac{1}{\sqrt{ab}} 
  \,\delta(a-b)\ , \qquad \text{for } a,b>0
\end{displaymath} 
gives the following orthogonality and completeness relations 
\begin{eqnarray}\label{qm11} 
  \skal{\phi_{\alpha ;m,E}}{\phi_{\alpha ;m',E'}} &=& 
  \delta_{m,m'}\, \delta(E-E')   
  \\ \label{qm12} 
  \sum_{m=-\infty}^{+\infty} \int_{0}^{\infty}dE\, \phi_{\alpha ;m,E} 
  (r,\theta) \, \overline{\phi_{\alpha;m,E}(r',E')}  & = & 
  \frac{1}{r}\,\delta(r-r')\, \delta(\theta-\theta') \ .  
\end{eqnarray} 
We will also need the integral kernel for the resolvent
$R_{0,\alpha}(z) := \left(H_0(\alpha)-z\right)^{-1}$ also called
Green's function, given by
\begin{equation}\label{qm13} 
  \matrele{r,\theta}{R_{0,\alpha}(k\pm i\varepsilon)}{r',\theta'} = 
   \pm\frac{i}{2} \sum_{m=-\infty}^{+\infty} e^{2im(\theta-\theta')} 
   J_{\mu}(kr_{<})  \,H_{\mu}^{\pm}(kr_{>}) \ .  
\end{equation} 
Here we have used the standard convention $r_{>}:=\max(r,r')$ and
$r_{<}:=\min(r,r')$.  To abbreviate notation we will sometimes make
the convention that $H_{\mu}^{\pm}$ denotes the first and second
Hankel function $H^{(1)}_{\mu}$ and $H^{(2)}_{\mu}$ respectively.
$I_{\mu}$ is the modified Bessel function and $K_{\mu}$ is the
MacDonald function (see e.g.\ \onlinecite{Watson22}). To establish
\eqref{qm13} one uses the well--known formula (see e.g.
\onlinecite{Magnus66}).
\begin{displaymath} 
  \label{qm14} 
  \int_{0}^{\infty}k\,dk\,\frac{J_{\mu}(kr)\, 
  J_{\mu}(kr')}{k^2+c^2} =I_{\mu}(cr_{<})\,K_{\mu}(cr_{>}) \quad
[\Ret\mu>-1,\Ret c>0] . 
\end{displaymath} 
Finally we give the kernel of of the unitary time evolution
$\exp\left(-itH_0(\alpha)\right)$:
\begin{eqnarray} 
  \label{qm15} 
  {\frak K}_{\alpha}(r,\theta;r',\theta';t) := \matrele{r,\theta}{ 
  e^{-itH_{0}(\alpha)}}{ r',\theta'}  
\end{eqnarray} 
We can make this explicit by adapting a calculation, which is quite
standard in the context of the Aharonov--Bohm effect
\cite{Aharonov59}. Along the lines of \onlinecite{Alvarez96,Gerbert89} we
obtain the following result:
\begin{equation} 
  \label{qm16} 
  {\frak K}_{\alpha}(r,\theta;r',\theta';t)= 
  {\frak K}_{\alpha,0}(r,\theta;r',\theta';t) 
  +\hat{\frak K}_{\alpha}(r,\theta;r',\theta';t) \ . 
\end{equation} 
Here the two subexpressions ${\frak K}_{\alpha,0}$ and $\hat{\frak
  K}_\alpha$ are given as
\begin{align} 
  \label{qm17} 
  {\frak K}_{\alpha,0}(r,\theta;r',\theta';t) =& \frac{1}{2\pi it}\, 
  e^{-\frac{1}{4it}(r^2+{r'}^2)} \,e^{i\alpha (\theta - 
    \theta'-\frac{\pi}{2} \sgn(\theta-\theta'))}\times  
  \\ \nonumber  
  &\times \cos\left(-\frac{\alpha\pi}{2} \sgn(\theta -\theta') 
  +\frac{rr'}{2t} \cos (\theta -\theta')\right) 
  \\ \label{qm18} 
  \rule{0mm}{7mm} \hat{\frak K}_{\alpha}(r,\theta;r',\theta';t) =& 
  \frac{i}{2\pi t}\cdot\frac{\sin (\pi\alpha)}{\pi}\, e^{-\frac{1}{4it}(r^2 
    +{r'}^2)}\times I_{\alpha}(\tfrac{rr'}{2t},\theta -\theta') \ , 
\end{align} 
where we have introduced 
\begin{equation} 
  \label{qm19} 
  I_{\alpha}(\rho,\chi)=\int_{-\infty}^{+\infty}dy\, e^{i\rho\cosh y} 
  \, \frac{e^{-y\alpha}}{1-e^{-2y-2i\chi}} \ . 
\end{equation}                            
The formal relation to the Aharonov--Bohm effect used in the derivation of
\eqref{qm17} and \eqref{qm18} does not come by accident. In fact, the
Hamiltonian of the two--anyon system coincides with the Hamiltonian of
the Aharonov--Bohm effect when restricted to the subspace of symmetric
wave functions. As has been realized before (see e.g.
\onlinecite{Wilczek90}) one can exploit this by describing the anyonic
dynamics with the help of Aharonov--Bohm scattering and thus
demonstrating the non--trivial character of the statistical
interaction. The description of the Aharonov--Bohm effect in terms of
scattering theory was first given by Aharonov and Bohm themselves
and later taken up by several authors, among others 
\onlinecite{Henneberger81,Ruijsenaars83,Sakoda97}. We will mostly follow 
the discussion presented in \onlinecite{Ruijsenaars83} because there the
time--dependent as well as the time-independent scattering formalism
are considered.
The wave operators and the scattering operator of the Aharonov--Bohm
effect are formally defined by
\begin{equation}\label{qm20} 
  \Omega^\pm_{AB}:=\slim_{t\to\pm\infty}e^{itH_0(\alpha)}e^{-itH_0} 
  \qquad\text{and}\qquad S_{AB}:=(\Omega^+_{AB})^\ast\Omega^-_{AB}
\end{equation} 
respectively. Here and henceforth $H_0$ denotes the bosonic
Hamiltonian $H_0(\alpha=0)$ and the symbol $\slim$ stands for the
strong operator limit. Furthermore, by writing \eqref{qm20} we
have implied the restriction of the Aharonov--Bohm scattering to the
subspace of symmetric functions or equivalently to the space $\cL^2(\cone)$. It has
been shown \cite{Ruijsenaars83} that the wave operators exist and are
complete, whence the scattering operator is unitary. For later use we
give the explicit form of the integral kernels of $\Omega^\pm_{AB}$,
i.e. the stationary scattering states, which can be obtained by
symmetrizing the results given in \onlinecite{Ruijsenaars83,Sakoda97},
\begin{equation}\label{qm21} 
\matrele{r,\theta}{\Omega^\pm_{AB}}{k,\theta'} 
=\frac{1}{\pi}\sum_{m\in\Z}(\pm i)^{|2m+\alpha|}J_{|2m+\alpha|}(kr) 
e^{i\,2m(\theta-\theta')}, 
\end{equation}   
where $\ket{k,\theta'}$ denotes the symmetric plane wave, i.e.  
\begin{displaymath}
  \skal{r,\theta}{k,\theta'}=\tfrac{1}{\pi}
  \cos(kr\cos(\theta-\theta')) \ .
\end{displaymath}
The scattering phase shift corresponding to $S_{AB}$ was first derived
by Henneberger \cite{Henneberger81}. Using his result we can rewrite the
scattering operator in the compact form
\begin{equation}\label{qm22} 
S_{AB}=e^{i\pi\alpha}P_\geq+e^{-i\pi\alpha}P_< 
\end{equation} 
with $P_\geq$ and $P_<$ denoting the spectral projections onto the
subspaces of positive respectively negative angular momentum. Equation
\eqref{qm22} serves physical intuition by clarifying the effect of the
gauge potential $A_\alpha$ defined in \eqref{qm0}. \\
However, of practical importance is the integral kernel of the
$S$--matrix in momentum space which was first given in
\onlinecite{Ruijsenaars83}. After symmetrization we obtain for $\alpha\in[0,1]$
\begin{equation}\label{qm23} 
\matrele{k,\theta}{S_{AB}}{k',\theta'}=4\delta(k^2-{k'}^2) 
\left[\cos\pi\alpha\,\delta(\Theta)+\frac{\sin\pi\alpha}{2\pi i}\, 
PV\left(1+i\cot\Theta\right)\right]. 
\end{equation}  
Here, $\Theta=\theta-\theta'$ and $PV$ stands for the principal value
prescription. From \eqref{qm23} one immediately derives the scattering
amplitude
\begin{eqnarray}\label{qm24} 
  f_{AB}(k,\Theta) &=& \left(\tfrac{\pi}{ik}\right)^{\frac{1}{2}} 
  \matrele{\theta}{T_{AB}(E=k^2)}{\theta'}
  \\ \nonumber 
  &=&\left(\tfrac{\pi}{ik}\right)^{\frac{1}{2}} 
  2\left[(\cos\pi\alpha-1)\,\delta(\Theta)+\frac{\sin\pi\alpha}{2\pi i}\, 
  PV\left(1+i\cot\Theta\right)\right], 
\end{eqnarray} 
where $\matrele{\theta}{T_{AB}(E)}{\theta'}$ denotes the on-shell  
matrix element of $T_{AB}:=S_{AB}-\id$ defined by the equation  
\begin{equation*} 
\matrele{k,\theta}{T_{AB}}{k',\theta'}= 
2\delta(E-E')\matrele{\theta}{T_{AB}(E)}{\theta'} 
\end{equation*} 
with $E=k^2$ and $E'={k'}^2$.  Note, that equation \eqref{qm24}
differs from what one would get when symmetrizing the amplitude
in the way calculated by Aharonov and Bohm \cite{Aharonov59}.  In their result
the contribution of the $\delta$--function was left out. This violates
the unitarity of the $S$--matrix as was pointed out by Ruijsenaars
\cite{Ruijsenaars83}. Thus, away from the forward direction we end up
with the following expression for the differential cross--section
\begin{eqnarray}\label{qm25} 
\frac{d\sigma_{AB}}{d\theta} &=& |f_{AB}(k,\theta)|^2\notag\\ 
&=& \frac{\sin^2\pi\alpha}{\pi k}\left(1+\cot^2\theta\right),
\quad\theta\neq 0, 
\end{eqnarray} 
which has a nontrivial angular dependence. Note that the total
cross--section is infinite if $\alpha\not\in\Z$ because of the
singular contribution in the forward direction as displayed in
\eqref{qm24}. The latter has been interpreted to be characteristic of
anyon statistics reflecting the long range nature of the statistical
interaction \cite[p23]{Wilczek90}. For $\alpha=0,1$, i.e. for bosons
and fermions, the cross section vanishes. This is obvious from
equation \eqref{qm22} which shows that bosons are not scattered at all
while fermions pick up a factor minus one.

\section{Potential Scattering} \label{sec:pot}
In this section we discuss two--particle potential scattering for
anyons. Time dependent scattering theory involves the comparison
between two dynamics, one of which is considered to be ``free'' in an
appropriate sense. Here, we shall consider the ``free'' dynamics to be
the one defined by $H_0(\alpha)$. This appears to be a natural
choice, if one views statistics to be an inherent property of
the particles. This is a generalization of the fermionic picture,
where usually the free time evolution acts on the space of \em
antisymmetric \em functions. Note, however, that this ``free'' dynamics
contains an interaction given by the long range gauge forces
encoding the statistics. In particular, non--integer values of
$\alpha$ lead to a highly non--trivial ``free'' time evolution similar
to the Aharonov--Bohm effect (see our discussion at the end of
section \ref{sec:qm}). As a consequence not all methods used in
potential scattering theory may be applied. For example, we do not
know how two adopt Enss'~\cite{Enss78} method, which makes use of
Fourier transformation and is therefore not suitable in the present
context.  Thus, in order to prove the existence of the wave operators and
the unitarity of the $S$--matrix, we will rely on Cook's method (see
e.g.  \onlinecite{Reed79III}) and the Kuroda--Birman
theorem~\cite{Kuroda62} respectively.
\Absatz 
Following our discussion in section~\ref{sec:qm} we will formulate the
two particle scattering theory in the center of mass system using the
Hilbert space $\cL^2 (\cone)$ with $H_{0}(\alpha) = -\Delta_{\alpha}$
being the free Hamiltonian. Let $V$, the potential, be a measurable
function on $\cone$. For example $V$ may result from a function, also
denoted by $V(z)$, on $\C$ with $V(-z)=V(z)$ and which acts as a
multiplication operator. We set
\begin{displaymath}\label{def1}
  H(\alpha) := H_{0}(\alpha )+V=-\Delta_{\alpha}+V
\end{displaymath}
as an operator on $\cL^2(\cone)$. At this point we are not concerned
with giving a criterium for self--adjointness of $H_0(\alpha)$; below
we specify a certain class of \em spherically symmetric \em potentials
for which this operator is self--adjoint.  The wave operators
$\Omega^\pm_\alpha$ for the pair $(H(\alpha ),H_{0}(\alpha))$ are
defined as follows
\begin{displaymath}\label{def2}
  \Omega^{\pm}_{\alpha}
  :=\slim_{t\to\pm\infty}\,e^{itH(\alpha )}\,e^{-itH_{0}(\alpha )}
\end{displaymath}
provided the strong operator limit, denoted by $\slim$, exists. Note
that in the present context the absolute continuous spectrum of
$H_{0}(\alpha)$ is the positive real axis including the origin and the
associated space is all of $\cL^2(\cone)$. The $S$--matrix is then
defined as
\begin{displaymath}\label{def3}
  S_{\alpha}:=(\Omega^{+}_{\alpha})^\ast \Omega^{-}_{\alpha}
\end{displaymath}
Now we have the
\begin{theo}
  \label{theo:pot:1}
  Let (1+r)V be in $\cL^2(\cone)$.  Then the wave operators
  $\Omega^{\pm}_{\alpha}$ for the pair $(H(\alpha ),H_{0}(\alpha))$
  exist for all $\alpha$.
\end{theo}
According to Cook's theorem (see e.g.~\onlinecite{Reed79III}) it suffices to
prove the following lemma.
\begin{lem}\label{lem:pot:1} 
  Under the conditions on V stated in the Theorem one has
  \begin{displaymath}
    \int_{t}^{\infty}\norm{V\,e^{-isH_{0}(\alpha)}\,\phi}\, ds
    <\infty
  \end{displaymath}
  for a dense set of $\phi$'s and for some $t=t(\phi)<\infty$.
\end{lem}
We will provide two proofs of this lemma. The first one will use the
explicit form of the integral kernel of $e^{-itH_{0}(\alpha)}$ given
in expressions \eqref{qm17}, \eqref{qm18} and \eqref{qm19}. The second
proof will make use of the asymptotic form of Bessel functions near
the origin and at infinity.
\begin{proof}[1.~Proof of lemma~\ref{lem:pot:1}:]
  It suffices to consider the case $\alpha\in (0,1)$ since the cases
  $\alpha =0$ and $\alpha=1$ are the already well known bosonic and fermionic
  situation respectively. We will
  choose $\phi$ to be of the form $r\phi
  (r,\theta)=\frac{\partial}{\partial r}\Phi(r,\theta)$ with $\Phi\in
  C^{\infty}_{0}(\cone)$. It is easy to see that such
  $\phi$'s form a dense set.\\
  Using \eqref{qm16}, we write
  \begin{displaymath}
    \modulus{(e^{-itH(\alpha)}\phi )(r,\theta )}\le\,F_{\alpha ,0}
    (r,\theta ;t) + \hat{F}_{\alpha}(r,\theta ;t) 
  \end{displaymath}
  with
  \begin{displaymath}
    F_{\alpha ,0}(r,\theta ;t)\,=\modulus{\int^{\infty}_{0}\int^{\pi}_{0}
      {\mathfrak K}_{\alpha ,0}(r,\theta ;r',\theta';t)
      \,\phi (r',\theta')\,r'dr'd\theta'}  
  \end{displaymath}
  \begin{displaymath}
    \hat{F}_{\alpha}(r,\theta ;t)\,=\,\modulus{\int^{\infty}_{0}\int^{\pi}_{0}
    \hat{{\mathfrak K}}_{\alpha}(r,\theta ;r',\theta';t)
    \,\phi (r',\theta')\,r'dr'd\theta'}.
  \end{displaymath}
  We start with an estimate of $F_{\alpha ,0}$. Partial integration
  w.r.t $r'$ gives
  \begin{displaymath}
    F_{\alpha ,0}(r,\theta ;t)\,\le\,\frac{1}{t^2}\int^{\infty}_{0}\int^{\pi}_{0}
    G_{0}(r,\theta ;r',\theta';t)\modulus{\Phi(r',\theta')}
    \,dr'd\theta'
  \end{displaymath}
  where $G_{0}$ satisfies an estimate of the form
  \begin{displaymath}
    0\le\ G_{0}(r,\theta ;r',\theta';t)\,\le\,(1+r)\,C
  \end{displaymath}
  uniformly in $r,\theta ,t,\alpha\in (0,1)$ and $(r',\theta')$ in the support 
  of $\Phi$. This gives
  \begin{displaymath}
    F_{\alpha ,0}(r,\theta ;t)\,\le\,C\,\frac{(1+r)}{t^2}
  \end{displaymath}
  and hence
  \begin{equation}
    \label{pot1}
    \int^\infty_0\int^\pi_0\modulus{V(r,\theta )\, F_{\alpha,0}
      (r,\theta ;t)}^2 r\,dr\,d\theta \le \frac{C}{t^4}
    \int^\infty_0 \int^\pi_0 \modulus{(1+r) \, V(r,\theta )}^2
    r\,dr\,d\theta
  \end{equation}
  uniformly in $\alpha\in (0,1)$.  We turn to an estimate of
  $\hat{F}_{\alpha}$. By~\eqref{qm18}
  \begin{displaymath}
    0 \le \hat{F}_{\alpha} (r,\theta;t) \le \frac{C}{t} \cdot
    \modulus{ \int^\infty_0 \int^\pi_0
    e^{-\frac{1}{4it}(r^2+{r'}^2)}\, I_{\alpha}(\rho,\chi)\, \phi
    (r',\theta')\,r\,dr'\,d\theta'} 
  \end{displaymath}
  with the notation
  \begin{displaymath}
    \rho\,=\frac{rr'}{2t},\quad\chi=\theta-\theta'.
  \end{displaymath}
  Adding and subtracting $I_{\alpha}(0,\chi)$ gives
  \begin{align}
    \label{pot2}
    0 &\le \hat{F}_{\alpha}(r,\theta ;t) \\
    &\le \frac{C}{t}\int^\infty_0
    \int^\pi_0 \modulus{I_{\alpha}(\rho,\chi )-I_{\alpha}(0,\chi )
      }\modulus{\phi(r',\theta')}\, r'\,dr'\,d\theta'
    \nonumber \\
    &+\frac{C}{t}\modulus{ \int^\infty_0 \int^\pi_0 e^{-\frac{1}{4it}
    (r^2+{r'}^2)} \,I_{\alpha}(0,\chi )\,\frac{\partial}{\partial
    r'} \Phi(r',\theta')\,dr'd\theta'} \ .
  \end{align}
  We need the following lemma which will be proved in
  Appendix~\ref{sec:est}.
  \begin{lem} \label{lem:pot:2}
    The quantity $I_{\alpha}(\rho ,\chi )$ satisfies the 
    estimates
    \begin{align*}
      \modulus{I_{\alpha}(\rho ,\chi )} & \le C 
      \\
      \modulus{I_{\alpha}(\rho ,\chi )-I_{\alpha}(0,\chi )}
      &\le C(\epsilon )\,\rho^{\epsilon}
    \end{align*}
    uniformly in $\rho$ and $\chi$ for all $0\le\epsilon <\min (\alpha
    ,1-\alpha )$.
  \end{lem}
  We use this lemma combined with a partial integration w.r.t.\ $r'$
  in the second term of the r.h.s of \eqref{pot2} to obtain
  \begin{displaymath}
    \modulus{\hat{F}_{\alpha}(r,\theta;t)} \le\,C(\epsilon)
    \left( \frac{r^\epsilon}{t^{1+\epsilon}}+\frac{1}{t^2} \right) \ .
  \end{displaymath}
  Therefore, since $r^{\epsilon}\,\le 1+r$ we finally arrive at
  \begin{multline}
    \label{pot3}
    \int^{\infty}_{0}\int^{\pi}_{0}\modulus{V(r,\theta )}^2
    \modulus{\hat{F}_{\alpha }(r,\theta ;t)}^2 r\,dr\,d\theta \leq
    \\
    C(\epsilon )\left(\frac{1}{t^2} +\frac{1}{t^{1+\epsilon}}
    \right)^2 \int^\infty_0 \int^{\pi}_{0}(1+r)^2\modulus{V(r,\theta
      )}^2 r\,dr\,d\theta .
  \end{multline}
  Combining \eqref{pot1} and \eqref{pot3} shows that $\norm{V
    e^{-itH_{0}(\alpha )}}$ is integrable in t on the interval
  $[1,\infty )$ say, concluding the 1.~Proof of lemma~\ref{lem:pot:1}.
\end{proof}
\begin{proof}[2.~Proof of lemma~\ref{lem:pot:1}:]
  For fixed and given $\alpha\in (0,1)$ we use the spectral
  decomposition of $H_{0}(\alpha)$ to obtain an $\alpha$-dependent
  unitary equivalence
  \begin{equation}
    \label{pot6}
    \hat{V}_{\alpha}  :\,\cL^2(\cone)\, \to \, \hat\cL=
    \underset{m\in \Z}{\oplus} \,\cL^2(\R^+,dE)   
  \end{equation}
  defined by
  \begin{displaymath}
    \label{pot7}
    \psi\longmapsto\hat{\psi}\equiv\, \{ \hat{\psi}_{m}\}_{m\in \Z}
    \qquad , \qquad 
    \hat{\psi}_{m}(E)\,=\skal{\phi_{\alpha ;m,E}}{\psi} \ ,
  \end{displaymath}
  such that $H_{0}(\alpha )$ just turns into a multiplication operator
  $\widehat{H_{0}(\alpha )}$:
  \begin{displaymath}
    (\widehat{H_{0}(\alpha )\psi})_{m}(E)=
    (\widehat{H_{0}(\alpha )}\hat{\psi})_{m}(E)=E\,\hat{\psi}_{m}(E).
  \end{displaymath}
  Via this isomorphism we have
  \begin{multline}
    \label{pot4}
    \norm{V e^{-itH_{0}(\alpha )}}^2 = \\
    \sum_{m,m'\in \Z}\int^\infty_0 dE\int^\infty_0 dE'\,
    e^{-it(E'-E)}\,\overline{\hat{\psi}_{m}(E)}\,v^{m,m'}_{\alpha}(E,E')
    \,\hat{\psi}_{m'}(E') \ , 
  \end{multline}
  where
  \begin{displaymath}
    v^{m,m'}_{\alpha}(E,E') =
    \skal{\phi_{\alpha;m,E}}{V^2\phi_{\alpha ;m',E'}} \ . 
  \end{displaymath}
  We now choose the following dense subspace $\hat{\cal D} \subset
  {\cal D}(\widehat{H_{0}(\alpha)})$ in $\hat\cL$
  \begin{displaymath}
    \hat{\cal D}= \left\{ \hat{\psi}\in\hat\cL \ \big\vert \
      \hat{\psi}_{m} \in C^\infty_0 (\R^+) \text{ and }
      \hat{\psi}_{m}\equiv \,0 \text{ for almost every } m \right\} \ .
  \end{displaymath}
  Correspondingly we set ${\cal D}=(\hat{V}_{\alpha})^{-1}\hat{\cal
    D}$.  We need the following lemma, which will be proved in
  Appendix~\ref{sec:est}.
  \begin{lem} 
    \label{lem:pot:3}
    Let V satisfy the conditions of the Theorem. For arbitrary $m$ and
    $m'$ the functions $v^{m,m'}_{\alpha}(E,E')$ have measurable
    partial derivatives up to order 3 in $E$ and $E'$ on $(0,\infty)$,
    which are essentially bounded on compact sets.
  \end{lem}
  With this lemma at hand we now proceed as follows. We use the
  identity
  \begin{displaymath}
    e^{itE}\,=\,\left(\frac{1}{it}\frac{\partial}{\partial E}\right)^{n}
    \,e^{itE}
  \end{displaymath}
  to perform 3 partial integrations w.r.t.\ E in \eqref{pot4}. This
  gives for $\psi\in{\cal D}$
  \begin{align*}
    \norm{V e^{-itH_{0}(\alpha)}\,\psi}^2 &\le \frac{1}{t^3}
    \sum_{m,m'}\int dE\int dE'
    \modulus{\partial^3_E \left\{ \overline{\hat{\psi}_{m}(E)} \,
    v^{m,m'}_{\alpha}(E,E')\right\} 
    }\modulus{\hat{\psi}_{m'}(E')} \\ 
    & \le\,C\,\frac{1}{t^{3}}
  \end{align*}
  which again shows that $\norm{Ve^{-itH_{0}(\alpha)}\,\psi}$ is
  integrable in $t$ in the interval $[1,\infty)$ say. This concludes
  the second proof of lemma~\ref{lem:pot:1}.
\end{proof}
Now we turn to the situation, where the potential is centrally
symmetric, i.e. where $V=V(r)$. According to \onlinecite{Bulla85} it is
possible to define $H(\alpha)=H_0(\alpha)+V$ as a self--adjoint operator,
if the potential is of the form
$\tfrac{\gamma}{r}+\tfrac{\beta}{r^b}+W(r)$ say, where $\gamma$ and
$\beta$ are arbitrary reals, $b\in[0,2)$ and $W$ is a bounded function
on $\R^+$. In general  $H(\alpha)$ is obviously diagonal w.r.t.\ the
decomposition~\eqref{qm3} such that
\begin{displaymath}
  H(\alpha )=\underset{m\in\Z}{\oplus} \,H_{2m}(\alpha)
  \quad\text{with}\quad
  H_{2m}(\alpha)\,=\,H_{0,2m}(\alpha )+V
\end{displaymath}
acting on $\cL^2(\R^+,rdr)$. This leads to a corresponding
decomposition for the wave operators and the $S$--matrix
\begin{equation}\label{pot00}
  \Omega^{\pm}_{\alpha}\,=\,\underset{m\in\Z}{\oplus}
  \,\Omega^{\pm}_{\alpha ,2m},
  \quad\text{where}\quad
  \Omega^{\pm}_{\alpha ,2m}\,=\slim_{t\to\pm\infty}\,e^{itH_{2m}(\alpha )}\, 
  e^{-itH_{0,2m}(\alpha)}
\end{equation}
and
\begin{displaymath}
  S_{\alpha}=\underset{m\in\Z}{\oplus} \,S_{\alpha ,2m}
  \quad\text{with}\quad
  S_{\alpha ,2m}\,=\,(\Omega^{+}_{\alpha ,2m})^\ast \,
  \Omega^{-}_{\alpha ,2m}.
\end{displaymath}
The Kuroda--Birman theorem~\cite{Kuroda62} now leads to the following
result.
\begin{theo}
  Let the centrally symmetric potential V satisfy
  \begin{displaymath}
    \int^{1}_{0}\modulus{V(r)}r\,dr\,+\int^{\infty}_{1}\modulus{V(r)} dr
    \, <\,\infty.
  \end{displaymath}
  Then the wave operators $\Omega^{\pm}_{\alpha ,2m}$ exist and are
  complete. In particular the $S$--matrices $S_{\alpha,2m}$ are
  unitary and so are the $S_{\alpha}$.
\end{theo}
We recall that the wave operators for an arbitrary pair $(H,H_{0})$
are called complete if they are unitary when considered as operators
from the absolutely continuous subspace of $H_{0}$ to the absolutely
continuous subspace of $H$. Note that these conditions on $V$ in the
central symmetric case are weaker than the conditions used in
theorem~\ref{theo:pot:1}.
\begin{proof}
  We have to show that the operator
  \begin{displaymath}
    \modulus V^{\frac{1}{2}}\,(\,H_{0,2m}(\alpha)\,+\,k^2)^{-1}
    \,V^{\frac{1}{2}}\nonumber
  \end{displaymath}
  in $\cL^2(\R^+,rdr)$ has finite trace class norm
  $\norm{\cdot}_{1}$, where $V^{\frac{1}{2}}(r)\,= \sgn(
  V(r))\cdot{\modulus{V(r)}^{\frac{1}{2}}.}$ Here it suffices to
  choose any $k^2\,>\,0$.  By \eqref{qm13} we have
  \begin{displaymath}
    \matrele{r}{ \left(H_{0,2m}(\alpha )\,+\,k^2 \right)^{-1}}{r'} =
    I_{\mu}(kr_{<})\, K_{\mu}(kr_{>})
  \end{displaymath}
  with $\mu=\modulus{2m+\alpha}$.  Therefore we obtain the following a
  priori estimate
  \begin{eqnarray}
    \label{pot5}
      \norm{\modulus V ^{\frac{1}{2}}
      \left(\,H_{0,2m}(\alpha)\,+\,k^2 \right)^{-1}
      \,V^{\frac{1}{2}}}_{1} 
    \le\,\int^\infty_0 \modulus{V(r)}\modulus{ I_{\mu}(kr)}
    \modulus{K_{\mu}(kr)} r\,dr \ .
  \end{eqnarray} 
  The following estimates for $I_\mu,K_\mu$ can be derived from the
  asymptotic behavior near the origin and at infinity (see e.g.
  \onlinecite{Watson22,Magnus66})
  \begin{align*}
      \text{for } kr \le 1 : \qquad 
      \modulus{I_{\mu}(kr)} & \le C(\mu)\,(kr)^{\mu} 
      && \modulus{K_{\mu}(kr)} \le\,C(\mu)\,(kr)^{-\mu}
      \\
      \text{for } kr \ge 1 : \qquad 
      \modulus{I_{\mu}(kr)} & \le C(\mu)\,\frac{e^{kr}}{\sqrt{kr}}
      && \modulus{K_{\mu}(kr)} \le C(\mu)\,\frac{e^{-kr}}{\sqrt{kr}}  
  \end{align*}
  Now we choose $k=1$. For this choice of $k$ the right hand side of
  \eqref{pot5} is bounded by
  \begin{displaymath}
    C(\mu) \left\{ \int^1_0\modulus{V(r)} r\,dr +
    \int^\infty_1\modulus{V(r)} dr \right\}
  \end{displaymath}
  concluding the proof of the theorem.
\end{proof}

\section{The Differential Cross--Section} 
\label{sec:cross} 
We now turn to the discussion of the two--particle scattering
cross--section.  The conventional approach when dealing with identical
particles is to compute first the cross--section for \em
distinguishable \em particles and in a second step to symmetrize or
anti--symmetrize it in order to describe boson or fermion scattering.
This is allowed because all observables commute with the projection
operators onto the subspaces of symmetric and antisymmetric
wave functions respectively (see e.g.\ \onlinecite{Taylor83} for a detailed account on
this issue). In the context of anyon statistics this procedure is not
applicable since no corresponding projection operators on the anyonic Hilbert spaces
are available. Hence, we will investigate the two--particle
cross--section in the center of mass system in a way similar to the
single particle case. Doing this one encounters an additional
difficulty, namely the presence of the gauge potential $A_\alpha$
encoding the statistics.
The latter gives rise to a non--trivial differential  
cross--section even if no {\em dynamical} potential is present, i.e. $V\equiv 0$.  
This should not come as a surprise since the ``free'' dynamics generated by  
$H_0(\alpha)$ is described by Aharonov--Bohm scattering in the infinite  
time limit (see our discussion in section \ref{sec:qm}). Thus, the  
cross--section will display both the statistical as well as the dynamical  
interaction represented by $A_\alpha$ and $V$ respectively.\\ 
\\ 
Let us first consider the simplest case where $\alpha=0$, i.e. the 
particles are bosons. Then the natural choice of a basis for describing 
the scattering is given by the {\em symmetric} plane waves $\ket{k,\theta}$, 
because the incoming asymptote is usually taken to have a sharply 
peaked momentum distribution. Moreover, we recall that the localization of  
some state $\psi$ at large times can be determined by means of its Fourier  
transform, 
\begin{equation}\label{cross1} 
\lim_{t\to\pm\infty}\int_\cC|e^{-itH_0}\psi|^2= 
\int_\cC d\theta\,k\,dk\,|\skal{k,\theta}{\psi}|^2, 
\end{equation} 
where $\cC:=\{(k,\theta):k\in\R_+,\;\theta\in (a,b)\subset[0,\pi]\}$ denotes  
a cone in real space on the left hand side and in momentum space on the right  
hand side. The above identity can be found in several text books on  
scattering theory, see e.g.\ \onlinecite{Pearson88}.\\  
If $\alpha\not\in\Z$ the flux--tube destroys 
translation invariance, whence the plane waves are not eigenstates of 
$H_0(\alpha)$. Thus, we have to look for an appropriate replacement such that  
the new basis diagonalizes $H_0(\alpha)$ and we still can form an incident  
wave packet with sharply peaked momentum distribution. It turns out that this  
can be achieved by making use of Aharonov--Bohm scattering theory. We assign  
to each symmetric plane wave denoted by $\ket{k,\theta}$ the corresponding  
stationary Aharonov--Bohm scattering state, that is  
\begin{equation}\label{cross2} 
\ket{k,\theta}^{in}:=\Omega^-_{AB}\ket{k,\theta}\quad\text{and}\quad 
\ket{k,\theta}^{out}:=\Omega^+_{AB}\ket{k,\theta} 
\end{equation} 
The symbols $\Omega^\pm_{AB}$ stand for the Aharonov--Bohm wave operators  
as defined in \eqref{qm20}. Note 
that each of the sets $\{\ket{k,\theta}^{in}\}$ and 
$\{\ket{k,\theta}^{out}\}$ forms a complete orthonormal system since 
the Aharonov--Bohm wave operators are unitary \cite{Ruijsenaars83}. Moreover,  
the basis elements are (improper) eigenstates of $H_0(\alpha)$ due to the  
intertwining relation  
\begin{displaymath}\label{cross3} 
H_0(\alpha)\Omega^\pm_{AB}=\Omega^\pm_{AB}H_0. 
\end{displaymath} 
Their explicit form was given in section \ref{sec:qm} equation \eqref{qm21}.  
Now, by construction every wave packet in the 
$\ket{k,\theta}^{\substack{in\\out}}$ basis under the 
``free'' anyonic time evolution $\exp(-itH_0(\alpha))$ approaches the 
corresponding wave packet in the plane wave basis $\ket{k,\theta}$ as 
$t\to\mp\infty$. Therefore, we shall refer to 
$\ket{k,\theta}^{\substack{in\\out}}$ as the anyonic state with 
incoming respectively outgoing (relative) momentum $(k,\theta)$. Note 
that according to their definition the ``in'' states are transformed 
into the ``out'' states by the Aharonov--Bohm $S$--matrix 
\begin{displaymath}\label{cross4} 
{}^{out}\!\skal{k,\theta}{k',\theta'}^{in}= 
\matrele{k,\theta}{S_{AB}}{k',\theta'}. 
\end{displaymath} 
An additional argument for the usefulness of the two bases  
$\ket{k,\theta}^{\substack{in\\out}}$ is the following generalization 
of Dollard's theorem to anyonic dynamics, 
\begin{equation}\label{cross5} 
\lim_{t\to\pm\infty}\int_\cC|e^{-itH_0(\alpha)}\psi|^2= 
\int_\cC d\theta\,k\,dk\, |{}^{\substack{in\\out}}\!\skal{k,\theta}{\psi}|^2. 
\end{equation} 
The proof is immediate. By construction the state  
$e^{-itH_0}(\Omega^\pm_{AB})^\ast\psi$ approaches the anyonic state  
$e^{-itH_0(\alpha)}\psi$ in norm as $t\to\pm\infty$.  
Applying \eqref{cross1} yields the desired relation.\\ 
We are now prepared to consider a single scattering event of two anyons 
when a potential $V$ is present. Denote now by $\psi$ the incoming wave 
function. According to \eqref{cross5} the probability that the particles are scattered 
into $\cC$ is given by  
\begin{eqnarray}\label{cross6} 
\lim_{t\to\infty}\int_\cC|e^{-itH_0(\alpha)}S_\alpha\psi|^2 
&=& \int_\cC d\theta k\,dk\,|{}^{out}\!\skal{k,\theta}{S_\alpha\psi}|^2, 
\end{eqnarray} 
where $S_\alpha$ is the scattering operator introduced in section  
\ref{sec:pot}. On physical grounds $\psi$ is assumed to have a sharply peaked 
momentum distribution at $t=-\infty$, whence it will be given in the basis  
$\ket{k,\theta}^{in}$. Therefore, we perform the following transformation 
\begin{eqnarray}\label{cross7} 
{}^{out}\!\skal{k,\theta}{S_\alpha\psi} &=& \int d\theta' k'dk'\, 
{}^{out}\!\matrele{k,\theta}{S_\alpha}{k',\theta'}^{in}\; 
{}^{in}\!\skal{k',\theta'}{\psi}\notag\\ 
&=&\int d\theta' k'dk'\,  
\matrele{k,\theta}{(\Omega^+_{AB})^\ast S_\alpha\Omega^-_{AB}}{k',\theta'}\;  
{}^{in}\!\skal{k',\theta'}{\psi}, 
\end{eqnarray}  
where in the second step we have used the defining relation 
\eqref{cross2}. Thus, we are lead to consider the operator  
\begin{displaymath}\label{cross8} 
S^{tot}_\alpha:=(\Omega^+_{AB})^\ast S_\alpha\Omega^-_{AB}, 
\end{displaymath} 
which can be identified with the $S$--matrix resulting from the wave
operators
\begin{displaymath}\label{cross9} 
\Omega^\pm(H(\alpha),H_0)=\slim_{t\to\pm\infty}e^{itH(\alpha)}e^{-itH_0}. 
\end{displaymath} 
This can be easily verified by using the well known chain rule for   
wave operators (see e.g.\ \onlinecite{Reed79III}) 
\begin{eqnarray}\label{cross10} 
\Omega^\pm(H(\alpha),H_0)&=&\Omega^\pm(H(\alpha),H_0(\alpha)) 
\Omega^\pm(H_0(\alpha),H_0)\notag\\ 
&=& \Omega^\pm_\alpha\;\Omega^\pm_{AB}. 
\end{eqnarray} 
Notice that the scattering operator $S^{tot}_\alpha$ incorporates the  
statistical as well as the dynamical interaction while $S_\alpha$ only 
takes care of the latter. This is best displayed by setting the 
potential equal to zero, 
\begin{displaymath}\label{cross11} 
V\equiv 0\quad\Rightarrow\quad S_\alpha=\id,\;S^{tot}_\alpha=S_{AB}. 
\end{displaymath} 
Therefore, in order to accommodate the anyon statistics of the particles we  
shall define the scattering amplitude $f_\alpha$ refering to the dynamical  
interaction $V$ as follows. Denote by $T^{tot}_\alpha$ the operator  
\begin{displaymath}\label{cross12} 
T^{tot}_\alpha:=(\Omega^+_{AB})^\ast (S_\alpha-\id)\Omega^-_{AB}= 
S^{tot}_\alpha-S_{AB}. 
\end{displaymath} 
Then the scattering amplitude is given by  
\begin{equation}\label{cross13} 
f_\alpha(k,\theta-\theta') := \left(\tfrac{\pi}{ik}\right)^{\frac{1}{2}} 
\matrele{\theta}{T^{tot}_\alpha(E=k^2)}{\theta'}, 
\end{equation} 
where $\matrele{\theta}{T^{tot}_\alpha(E)}{\theta'}$ denotes the on--shell  
matrix element defined by the relation  
\begin{displaymath}\label{cross14} 
\matrele{k,\theta}{T^{tot}_\alpha}{k',\theta'}= 
2\delta(E-E')\matrele{\theta}{T^{tot}_\alpha(E)}{\theta'} 
\end{displaymath} 
with $E=k^2$ and $E'={k'}^2$. From \eqref{cross6} and \eqref{cross7} one can now  
derive the differential cross--section analogously to the single-particle  
case, e.g.\ \onlinecite{Taylor83,Pearson88}. Away from the forward direction one  
ends up with the expression  
\begin{eqnarray}\label{cross15} 
\frac{d\sigma}{d\theta}(E_0,\theta_0)&=& \pi E_0^{-\frac{1}{2}} 
\modulus{\matrele{\theta}{T^{tot}_\alpha(E_0)+T_{AB}(E_0)}{\theta_0}}^2
\notag\\ 
&=&  |f_\alpha(k_0,\theta-\theta_0)+f_{AB}(k_0,\theta-\theta_0)|^2 
\end{eqnarray} 
with $\theta\neq\theta_0$. Here, $E_0=k_0^2,\theta_0$ determine energy
and direction of the incident particle beam and
$\matrele{\theta}{T_{AB}(E)}{\theta'}$ denotes the on-shell matrix
element of the operator $T_{AB}=S_{AB}-\id$ (see section \ref{sec:qm}
equation \eqref{qm25} for the explicit expression of the
Aharonov--Bohm scattering amplitude $f_{AB}$). Note that we have a
normalization factor $\pi$ instead of $2\pi$ in \eqref{cross15}
because the polar angle is restricted to the interval $[0,\pi]$. As 
mentioned in section \ref{sec:qm} the total Aharonov--Bohm
cross--section is infinite for
$\alpha\not\in\Z$, whence the total cross--section corresponding to
\eqref{cross15} is infinite as well.  However, the above differential
cross--section coincides with the usual one for bosons or fermions if
we set $\alpha=0$ and $\alpha=1$ respectively. This can be most easily
seen by use of the defining relation \eqref{cross2} for the ``in'' and
``out'' states which for the values $\alpha=0,1$ become symmetric
and antisymmetric plane waves respectively (compare
\eqref{qm21}). In case $\alpha=1$, however, this is only true
  up to an angular dependent phase factor which comes in by
  representing fermions as bosons with attached flux--tubes. This does
  not influence the outcome since the differential cross--section is
  given by the square modulus of the scattering amplitude. Thus, as special cases 
we obtain the familiar result that the differential cross--section for bosons and fermions is
given by the square modulus of the symmetrized and 
anti--symmetrized scattering amplitude respectively. In particular,
$f_{AB}$ vanishes and the total cross--section becomes finite for
suitable short range interactions $V$.
 
\section{Jost Functions and Levinson's Theorem} \label{sec:Jost}
In this section we will introduce Jost functions \cite{Jost47} indexed by a
continuous angular momentum and discuss their properties. This
continuous parameter leads to an alternative formulation of the
generalized Levinson theorem. We start by adapting the standard theory
and results of Jost functions to the present situation (see e.g.
\onlinecite{Newton60,Newton86,Alfaro65,Taylor83,Reed79III}). Conditions on
the spherically symmetric potential will be presented at the
appropriate places.
\\
Consider the unitary map $U:\cL^2(\R^+,rdr)\to\cL^2(\R^+,dr)$ given by
\begin{equation}
  \label{jost1}
  \psi(r)\longmapsto \sqrt{\,r}\,\psi (r) \ .
\end{equation}
Under this map the Hamiltonian 
$H_{2m}(\alpha )=H_{0,2m}(\alpha)+V$ turns into the operator
\begin{equation}\label{jost2}
  h_\mu\,=\,h_{0,\mu}\,+\,V
\end{equation}
with $\mu\,=\,\mid 2m+\alpha\mid\,>\,0$, where
\begin{equation}
  \label{jost3}
  h_{0,\mu}\,=\,-\left(\frac{d^2}{dr^2}\,-\,\frac{\mu^2-\frac{1}{4}}{r^2}\right)
  \ .
\end{equation}
In this section $\mu$ will be allowed to be any positive number. Also
in this section the notation $\phi_0(r;k,\mu )$ will be reserved for
the regular solutions (so called because of their behavior near
$r=0$) of the free Schr\"odinger equation $(h_{0,\mu}-k^{2})\,\psi=0$
given as
\begin{equation}
  \label{jost4}
  \phi_0(r;k,\mu)\,=\,\sqrt{\,\tfrac{\pi kr}{2}}\,J_\mu(kr) \ .
\end{equation}
With the help of these solutions the orthogonality and completeness
relations \eqref{qm11} and \eqref{qm12} are now reformulated as
\begin{subequations}
  \begin{align}\label{jost5}
    \int_{0}^{\infty}\phi_{0}(r;k,\mu)\;\phi_{0}(r';k,\mu)\,dk\,
      &=\,\frac{\pi}{2}\,\delta (r-r')\\
    \label{jost6}
    \int_{0}^{\infty}\phi_{0}(r;k,\mu)\;\phi_{0}(r;k',\mu)\,dr\,
    &=\,\frac{\pi}{2}\,\delta (k-k') \ .    
  \end{align}
\end{subequations}
We will also need the irregular, free solutions $\chi^{\pm}_{0}$ given as
\begin{displaymath}
  \label{jost7}
  \chi^{\pm}_{0}(r;k,\mu )\,=\,\pm i\,\sqrt{\,\tfrac{\pi kr}{2}}\,
  H^{\pm}_{\mu}(kr) \ ,
\end{displaymath}
With help of these solutions the free Green's functions read
(compare \eqref{qm13})
\begin{equation}
  \label{jost8}
  G^\pm_0(r,r';k,\mu )=
  \matrele{r}{(h_{0,\mu}-k^2\mp i\epsilon)^{-1}}{r'}
  = \tfrac{1}{k}\,\phi_0(r_<;k,\mu )\; \chi^\pm_0(r_>;k,\mu ) \ .
\end{equation}
\begin{defi}
  For given $V$ the function $\phi\,=\,\phi(r;k,\mu )$ is 
  the regular solution of the equation 
  $(h_{\mu}-k^{2})\,\psi\,=\,0$ which for $r\to 0$ approximates the free,
  regular solution $\phi_0$:
  \begin{displaymath}\label{jost9}
    \lim_{r\to 0}\;\left(\frac{2}{kr}\right)^{\mu +\frac{1}{2}}\,
    \frac{\Gamma(\mu +1)}{\sqrt{\pi}}\,\phi(r;k,\mu )\,=\,1 \ .
  \end{displaymath}
\end{defi}
As in the standard theory (see e.g.\ \onlinecite{Newton60,Newton86}) one
establishes the following facts.
\begin{itemize}
\item[(A0)] $\phi$ is the solution to the integral equation
  \begin{equation}\label{jost10}
    \psi\,=\,\phi_0\,-\,G^<_0\,V\,\psi\quad\text{with}\quad 
    G^<_0(r,r';k,\mu)=\Theta (r-r')\,g_{0}(r,r';k,\mu ) \ .
  \end{equation}
  Here $\Theta$ denotes the Heaviside step function and $g_0$ is given as
  \begin{equation}
    \label{jost11}
    g_{0}(r,r';k,\mu )= \tfrac{1}{k}[\phi_0(r;k,\mu )\,\chi^-_0(r';k,\mu ) 
    -\,\chi^-_0(r;k,\mu)\,\phi_0(r';k,\mu )\,] \ .
  \end{equation}
\item[(B0)] If the potential $V$ satisfies the condition
  \begin{displaymath}\label{jost12}
    \int^\infty_0 rdr\,|V(r)|;<\;\infty
  \end{displaymath}
  then the regular solution $\phi$ to the eigenvalue problem
  $(h_{0,\mu}\,+\,\lambda V\,-k^2)\,\psi\,=0$ has a convergent power 
  series expansion in $\lambda$ of the form
  \begin{equation}
    \label{jost13}
    \phi\,=\,\sum_{n=0}^\infty\,\lambda^n\,\phi_n\quad\text{with}\quad
    \phi_{n=0}=\phi_0\quad\text{and}\quad\phi_n=-G^<_0\,V\,\phi_{n-1},\,n\ge
    1\ .
  \end{equation}       
\item[(C0)] For fixed $\mu\,>\,0,\,r\,>\,0$ and real $V$ the solution $\phi$ 
  has an 
  analytic continuation in $k$ into the complex plane with a cut along the 
  negative imaginary axis. There one has the relation $(\,k\,>\,0)$
  \begin{equation}\label{jost14}
    \phi (r;-ik-0,\mu )\,=\,e^{i\pi (\mu+\frac{1}{2})}
    \;\phi (r;ik,\mu )\,=\,e^{2i\pi (\mu+\frac{1}{2})}
    \;\phi (r;-ik+0,\mu ) \ .
  \end{equation} 
\end{itemize}
The proofs of (A0), (B0) and (C0) are as in ordinary 3-dimensional
Schr\"odinger theory (see e.g.\ \onlinecite{Newton86}). As a byproduct of
the proof one also has the two estimates:
\begin{subequations}
  \begin{align}
    \label{jost15}
    \modulus{\phi_n(r;k,\mu )}\,<& \, e^{\modulus{\Im k}r}\,
    \left(\frac{\modulus k r}{1+\modulus k r}\right)^{\mu+\frac{1}{2}}
    \frac{C(\mu )^{n+1}}{n!}\left[\int_0^rdr'\,
      \frac{r'\modulus{V(r')}}{1+\modulus k r'}\right]^n
    \\ \label{jost16}
    \modulus{\phi (r;k,\mu )}\,<&\, C(\mu )e^{\modulus{\Im k}r}\;
    \left(\frac{\modulus k r}{1+\modulus k r}\right)^{\mu +\frac{1}{2}}
  \end{align}
\end{subequations}
We will prove these estimates in the Appendix~\ref{sec:est2}.
Relation \eqref{jost14} is a consequence of the relation
\begin{displaymath}\label{jost17}
  J_\mu(e^{in\pi})\;=\;e^{in\pi\mu}\;J_\mu(z) \ .
\end{displaymath}
We now introduce the irregular solutions of the eigenvalue equation
with potential as those solutions $\chi^\pm$ which approximate the
free irregular solutions $\chi^\pm_0$ at $r=\infty$.
\begin{defi}
  The functions $\chi^\pm\,=\,\chi^\pm(r;k,\mu )$ are the solutions
  of the eigenvalue problem $(\,h_{\mu}-k^{2})\,\psi=0$ satisfying the boundary
  conditions at infinity of the form
  \begin{displaymath}\label{jost18}
    \lim_{r\to\infty}\;e^{\mp i(kr-\frac{\pi}{2}\,\mu\,+\frac{\pi}{4})}\,
    \chi^{\pm}(r;k,\mu )\,=\,1 \ .
  \end{displaymath}
\end{defi}
These solutions will be called Jost solutions. With the help of these
solutions $\phi$ and $\chi^{\pm}$ the Green's function for the full
Hamiltonian $h_{\mu}$ now takes a form analogous to the free case (see
\eqref{jost8})
\begin{equation}\label{jost19}
  G^\pm(r,r';k,\mu )=\matrele{r}{(h_\mu-k^2\mp i\epsilon)^{-1}}{r'}
  =\frac{\phi (r_<;k,\mu )\;\chi^\pm(r_>;k,\mu)}{W(\chi^\pm,\phi)} \ ,
\end{equation}
where $W$ is the Wronskian, i.e. 
\begin{displaymath}\label{jost20}
  W(\psi_{1},\psi_{2})(r)\,=\,\psi_{1}(r)\,\partial_r\psi_{2}(r)\,
  -\,\psi_{2}(r)\,\partial_r\psi_{1}(r) \ .
\end{displaymath}
We recall that since $\phi$ and $\chi^{\pm}$ are solutions of the same
eigenvalue equation, the Wronskian in \eqref{jost19} is actually
independent of $r$, whence the notation.\\
Again one may establish the following facts (see e.g.\ \onlinecite{Newton86}).
\begin{itemize}
\item[(A$\pm$)] The functions $\chi^{\pm}$ are the solutions to the integral 
  equations
  \begin{equation}
    \label{jost21}
    \psi \, =\,\chi^\pm_0\,-\,G^>_0\,V\,\psi\ , \quad
    G^>_0(r,r';k,\mu):=\Theta (r'-r)\,g_{0}(r,r';k,\mu ) \ .
  \end{equation}
  Here, $g_0$ denotes the function defined in \eqref{jost11} which 
  can be rewritten as
  \begin{displaymath}\label{jost22}
    g_0(r,r';k,\mu )=\tfrac{i}{2k}\left[\chi^+_0(r;k,\mu)\chi^-_0(r';k,\mu)
      -\chi^-_0(r;k,\mu)\chi^+_0(r';k,\mu)\right] \ .
  \end{displaymath}          
\item[(B$\pm$)] If the potential $V$ satisfies the condition
  \begin{equation}\label{jost23}
    \int^{\infty}_{0}rdr\,(1+r)|V(r)|\;<\;\infty \ ,
  \end{equation}
  the irregular solutions $\chi^\pm$ to the eigenvalue problem
  $(h_{0,\mu}\,+\,\lambda V\,-k^2)\,\psi\,=0$ with $\pm\Im k\ge\,
  0,\,k\,\neq\,0$ have a convergent power series expansion in $\lambda$
  of the form
  \begin{displaymath}
    \label{jost24}
    \chi^\pm=\sum_{n=0}^\infty\lambda^n\chi^\pm_n\quad\text{with}
    \quad\chi^\pm_{n=0}=\chi^\pm_0\quad\text{and}\quad\chi^\pm_n=
    -G^>_0V\chi^\pm_{n-1},\,n\ge 1 \ .
  \end{displaymath}
\item[(C$\pm$)] For $V$ real and $\mu\,>\,0,\,r\,>\,0$ the functions
  $\chi^\pm$ are analytic in $k$ for $\pm\Im k\,>0$ with continuous
  extensions to the real axis
  except possibly for a singularity of order $\mu-\,\frac{1}{2}$ at $k=0$. 
  In case $V$ satisfies
  \begin{equation}\label{jost25}
    \int_{0}^{\infty}rdr\,| V(r)|\,e^{\eta r}\,<\,\infty,\qquad\eta>\,0
    \ ,
  \end{equation}
  then the functions $\chi^{\pm}$ have an analytic continuation into the
  domains given by $\pm\Im k\,>\,-\frac{\eta}{2}$ for all
  $\mu\,>\,0\,,r\,>\,0$ with the exception of a branch cut from $k=0$ to
  $k=\mp i\frac{\eta}{2}$.  There one has the relation $(\,k\,>\,0)$
  \begin{displaymath}\label{jost26}
    \chi^{\pm}(r;\mp (ik+0),\mu )\,
    =\,\chi^{\pm}(r;\mp (ik-0),\mu )\;+\;2i\cos \pi\mu\;
    \chi^{\pm}(r;\pm ik,\mu ) \ .
  \end{displaymath}
\end{itemize}
The proof of these statements is similar to the one in the regular
case and involves a proof of the bounds
\begin{multline}
  \label{jost27}
  \modulus{\chi^\pm_n(r;k,\mu )} < e^{\mp \Im(k)r}\,
  \left(\,\frac{\modulus k r}{1+\modulus k r}\right)^{
    -\mu+\frac{1}{2}}\times \\
  \times\,\frac{C(\mu )^{n+1}}{n!}\,  \left[\int_r^\infty dr'\,
    e^{\modulus{\Im k}r'\mp
      \Im(k)r}\frac{r'\modulus{V(r')}}{1+\modulus k r'}\right]^{n}  
\end{multline}
and
\begin{equation}
  \label{jost28}
  \modulus{\chi^{\pm}(r;k,\mu )} < C(\mu ) e^{\mp\Im(k)r}
  \left(\,\frac{\modulus k r}{1+\modulus k r}\right)^{-\mu+\frac{1}{2}}
\end{equation}
for all $\mu>0$ and $r>0$. We will establish these bounds in 
appendix~\ref{sec:est2}. Due to the uniqueness of the solutions
$\chi^\pm$ and the multivaluedness of the Hankel functions (see e.g.\ 
\onlinecite{Magnus66}) one has the relation
\begin{displaymath}
  \label{jost29}
  \overline{\chi^{\pm}(r;k,\mu )}\,=\,\chi^{\mp}(r;\overline{k},\mu)
  =\,\mp i\,e^{\pm i\pi \mu}\,\chi^{\pm}(r,-\overline{k},\mu) \ .
\end{displaymath}
We are now in the position to introduce the Jost functions in the
present context and establish their properties. Since $\chi^+$ and
$\chi^-$ form a basis of solutions, $\phi$ is a linear combination of
these two solutions. As in the standard theory (see e.g.\ 
\onlinecite{Newton86}) it turns out that this linear combination is of the
form
\begin{displaymath}
  \label{jost30}
  \phi (r;k,\mu )= \tfrac{i}{2}\left[F(k,\mu )\chi^-(r;k,\mu )
    -F(-k,\mu )\chi^+(r;k,\mu) \right] \ ,    
\end{displaymath}
where the coefficient function $F=F(k,\mu)$ is called the {\em Jost
  function} given by the Wronskian
\begin{equation}
  \label{jost31}
  F=\tfrac{1}{k} W(\chi^+,\phi) \ .
\end{equation}
The Jost function was first introduced by and named after R.~Jost.
Note, however, that our definition of $F$ follows the convention
introduced by R.G. Newton \cite{Newton86}, which differs from
the original one given in \onlinecite{Jost51}. We have the first main result
concerning these Jost functions.                                
\begin{theo}
  Let the potential V satisfy the condition \eqref{jost23}. 
  Then for fixed $\mu >0$ $F(k,\mu )$ extends to an analytic function in $k$ in 
  the open upper half plane, which is continuous up to the real axis with the 
  exception of the origin.\\ 
  In case the potential satisfies the stronger condition \eqref{jost25}, 
  $F(k,\mu )$ is analytic in $k$ in $\{\,k\,\mid\,\Im k\,>\,-\frac{\eta}{2}\}$ 
  except for a cut on the negative imaginary axis. Then $F$ also satisfies the 
  relation $(k\,>\,0)$
  \begin{displaymath}
    \label{jost32}
    F(-ik-0,\mu )\,=\,-e^{2i\pi\mu}F(-ik+0,\mu)\,+\,2\cos
    \pi\mu\:F(ik,\mu) \ .
  \end{displaymath}
\end{theo}
\noindent
The proof follows from the representation for $F$ 
\begin{equation}\label{jost33}
  F(k,\mu )\,=\,1+\frac{1}{k}\,
  \skal{\chi^{-}_{0}(\cdot\;;k,\mu )}{V\phi(\cdot\;;k,\mu)} \ ,
\end{equation}
the estimate \eqref{jost16} for $\phi$ and the estimate \eqref{jost28} which
is also valid for $\chi^{+}_{0}$.  To establish this relation we use
the integral equation \eqref{jost10} for $\phi$ which for large $r$ gives
\begin{multline} \label{jost34}
  \phi(r;k,\mu ) \xrightarrow{r\to\infty} \frac{i}{2} \, \Biggl[
  \chi^-_0(r;k,\mu)\left\{1+\frac{1}{k}\,
    \skal{\chi^-_0(\cdot\,;k,\mu )}{V\phi(\cdot\;;k,\mu
      )}\right\}
  \\  
  \quad -\chi^{+}_{0}(r;k,\mu )
  \left\{1+\frac{1}{k}\skal{\chi^+_0(\cdot\;;k,\mu )}{V\phi(\cdot\;;k,\mu)}\right\}
  \Biggr] \ . 
\end{multline}
Since $\chi^{\pm}(r;k,\mu)$ approaches $\chi^{\pm}_{0}(r;k,\mu )$ for
$r\to\infty$, relation \eqref{jost33} follows from this behavior.
\Absatz Note that making use of the analytic properties just proven
there is an equivalent definition of the Jost function in terms of
Fredholm theory, see e.g.\ \onlinecite{Newton86,Reed79II}. Consider the
operator $G^+_0(k,\mu)V$ with $G^+_0$ the free Green's operator and
$V$ satisfying relation \eqref{jost23}. Then for $k$ positive
imaginary ($k\in i\R^+$) the operator $G^+_0(k,\mu)V$ is trace class,
whence its Fredholm determinant $\det(\id+G^+_0(k,\mu)V)$ is well
defined. It turns out that the Jost function is just the analytic
continuation of this determinant to the upper half complex plane.                                
\Absatz
We note that the series expansion \eqref{jost13} for the regular solution
$\phi$ gives the series expansion
\begin{displaymath}\label{jost35}
  F\,=\,1+\frac{1}{k}\sum^\infty_0 F^n\quad\text{with}\quad
  F^n(k,\mu
  )\,=\,\skal{\chi^-_0(\cdot\;;k,\mu)}{V\phi^n(\cdot\;;k,\mu)}\ .
\end{displaymath}
The bound \eqref{jost15} for $\phi_n$ combined with the bound \eqref{jost28}
for $\chi^+_0$ gives the bound
\begin{displaymath}\label{jost36}
  \modulus{F^{n}(k,\mu )}<C(\mu )\frac{C^n}{n!}
  \int^\infty_0 dr\,e^{(\modulus{\Im k} -\Im k )r}
  \frac{\modulus k r}{1+\modulus k r}\modulus{V(r)} \ ,
\end{displaymath}  
while the bound \eqref{jost16} gives 
\begin{displaymath}\label{jost37}
  \mid F(k,\mu )-1\mid\,\le\,C(\mu )\int^{\infty}_{0}dr\,
  e^{(\mid\Im k\mid -\Im k)r}\,
  \frac{r|V(r)|}{1+|k|r} \ .
\end{displaymath}
The importance of the Jost function stems from its role in the
discussion of the $S$--matrix. Let $\Omega^{\pm}_{\mu}$ be the wave
operators for the pair $(h_{\mu},h_{0,\mu})$ such that one has (see
\eqref{pot00} and \eqref{jost1})
\begin{displaymath}\label{jost38}
  \Omega^{\pm}_{\mu}\,=\,U\,\Omega^{\pm}_{\alpha ,2m}\,U^{-1}\quad\text{and}\quad
  S_{\mu}\,=\,\left(\Omega^{+}_{\mu}\right)^\ast\,\Omega^{-}_{\mu}
  \,=\,U\,S_{\alpha,2m}\,U^{-1} \ .
\end{displaymath}
Since $S_{\mu}$ commutes with $h_{0,\mu}$ there is a
decomposition of $S_{\mu}$ in the form
\begin{displaymath}\label{jost39}
  S_{\mu}\,=\,\int_{\oplus}dk\,S(k,\mu )
\end{displaymath}
with respect to the spectral decomposition of $h_{0,\mu}$. Since the
spectrum of $h_{0,\mu}$ is not degenerate, $S(k,\mu )$ acts in a
one--dimensional Hilbert space and therefore is a complex number of
absolute value one, i.e.
\begin{displaymath}\label{jost40}
  S(k,\mu )\,=\,e^{2i\delta(k,\mu)}\;,k\ge 0\; ,\mu\ge 0
\end{displaymath}
with $\delta(k,\mu)$ being the phase shift. We define by 
\begin{displaymath}\label{jost41}
  f(k,\mu )\,:=\,\frac{e^{2i\delta (k,\mu )}-1}{\sqrt{\pi ik}}
\end{displaymath}
the partial wave amplitude for the angular momentum $\pm\mu$, such that 
we obtain the following partial wave decomposition of the scattering amplitude given in \eqref{cross13}
\begin{eqnarray}\label{jost41a}
f_\alpha(k,\theta-\theta') &=& (\tfrac{\pi}{ik})^\frac{1}{2}
\matrele{\theta}{T^{tot}_\alpha(k^2)}{\theta'}\notag\\
&=&(\tfrac{\pi}{ik})^\frac{1}{2}\,\tfrac{1}{\pi}\sum_{m\in\Z} 
e^{-i\pi|2m+\alpha|}(e^{2i\delta(k,|2m+\alpha|)}-1)\,e^{i2m(\theta-\theta')}\notag\\
&=&\sum_{m\in\Z}e^{-i\pi|2m+\alpha|}f(k,|2m+\alpha|)\,e^{i2m(\theta-\theta')}.  
\end{eqnarray}
Here we have used the relation 
$\matrele{k,\theta}{T^{tot}_\alpha}{k',\theta'}=
{}^{out}\!\matrele{k,\theta}{S_\alpha-\id}{k',\theta'}^{in}$ as well as the 
identities
\begin{eqnarray*}
\skal{r,\theta}{k,\theta'}^{\substack{in\\out}} &=& 
\sqrt{2\pi}\,\tfrac{1}{\pi}\sum_{m\in\Z}(\mp i)^{|2m+\alpha|}
\phi_{\alpha;m,E}(r,\theta)\,e^{-i2m\theta'}\\
\skal{\phi_{\alpha;m,E}}{S_\alpha\phi_{\alpha;m',E'}}&=&
\delta_{mm'}\delta(E-E')\,e^{2i\delta(k,|2m+\alpha|)}.
\end{eqnarray*}
Now set
\begin {displaymath}\label{jost42}
  \phi^{\pm}(r;k,\mu )\,=\,(\Omega^{\pm}_{\mu}\,\phi_{0})(r;k,\mu ) \ .
\end {displaymath}
By construction one has a solution of the eigenvalue equation 
$(h_{\mu}-k^{2})\psi\,=0$ and $\phi^{\pm}$ satisfies the
Lippmann-Schwinger equation
\begin {displaymath}\label{jost43}
  \phi^{\pm}(\cdot\;;k,\mu )\,=\,\phi_{0}(\cdot\;;k,\mu )
  -(h_{\mu}-k^{2}\pm i0)^{-1}\,V\,\phi^{\pm}(\cdot\;;k,\mu ) \ .
\end {displaymath}
With the help of $\phi^{\pm}$ one may express the 
partial wave amplitude as                                
\begin {displaymath}
  \label{jost44}
  f(k,\mu )\,=\,-\tfrac{2}{k^2}\,(\tfrac{ik}{\pi})^\frac{1}{2}\,
  \skal{\phi_{0}(\cdot\;;k,\mu)}{V\,\phi^{+}(\cdot\;;k,\mu)} \ .
\end {displaymath}
The solution $\phi^{+}$ of the Lippmann-Schwinger equation has the
following asymptotic behavior for $r\to\infty$.                               
\begin {displaymath}\label{jost45}
  \phi^+(r;k,\mu)\,\xrightarrow[r\to\infty]{}\,\phi_{0}(r;k,\mu )+
  \tfrac{1}{2}\sqrt{\pi k}\,f(k,\mu )\,e^{i(kr-\frac{\pi}{2}\mu)} \ ,
\end {displaymath}
where we have used 
$\chi^{+}_{0}(r;k,\mu )\,\xrightarrow[r\to\infty]{}\,
e^{i(kr-\frac{\pi}{2}\mu+\frac{\pi}{4})}$. Alternatively, using the relation 
$\phi_0\,=\,\tfrac{i}{2}\left(\,\chi^-_0
  -\chi^+_0\right)$ one gets 
\begin {displaymath}\label{jost46}
  \phi^{+}(r;k,\mu )\,\xrightarrow[r\to\infty]{}\,\frac{i}{2}
  \left(\,\chi^{-}_{0}(r;k,\mu )-e^{2i\delta (k,\mu
      )}\,\chi^{+}_{0}(r;k,\mu )\right) \ .
\end {displaymath}
With the same arguments as in the standard theory (see e.g.\ 
\onlinecite{Newton60,Newton86,Reed79III}) one proves
\begin{lem}
  Let $\mu\,>0$ be fixed and assume that V satisfies the condition 
  \eqref{jost23}. Then one has
  \begin{enumerate}
  \item[$(i)$] For $k$ on the real axis $(k\in \R^{+})$ the following
    identities are valid:
    \begin{align}
      \label{jost47}
      \phi(r;k,\mu )\,&=\,F(k,\mu )\,\phi^{+}(r;k,\mu )
      \\
      \label{jost48}
      S(k,\mu )\,=\,e^{2i\delta (k,\mu )}\,
      &=\frac{\overline{F}(k,\mu )}{F(k,\mu )}\,=\,\frac{F(-k,\mu )}{F(k,\mu )}
    \end{align}
    Furthermore, $F(k,\mu )$ can only vanish at the origin.\\
  \item[$(ii)$] The zeros of $F(k,\mu )$ in the upper half plane
    $\Im k \,>\,0$ all lie on the positive imaginary axis and are
    simple. In particular $k\in i\R^+$ is a zero of $F(k,\mu )$ iff
    $k^{2}$ is the energy of a bound
    state of $h_{\mu}$ with angular momentum $\pm\mu$.\\
  \item[$(iii)$] The following limit relation is valid in the closed
    upper half plane
    \begin {displaymath}\label{jost49}
      \lim_{k\to\infty}\,F(k,\mu )\,=\,1 \ .
    \end {displaymath}
  \end{enumerate}
\end{lem}
Since the proof is analogous to the one given in three dimensional
Schr\"odinger theory \cite{Newton60,Newton86,Reed79III} we omit it
here.  Now, the famous Levinson's theorem follows as a corollary by
use of the residue theorem. Only the so called resonance case, i.e.
$F(k=0,\mu)=0$, requires some more work.
\begin{theo}\label{theo:jost:2}
  Let $\mu>0$ and $V$ satisfy \eqref{jost23}. If $F(k=0,\mu)\neq 0$ one has 
  the following relation between the phase shift $\delta(k,\mu)$ and the 
  number $n_\mu$ of bound states with angular momentum $\pm\mu$, 
  \begin{equation}
    \label{jost50}
    \delta(k=0,\mu)-\delta(k=\infty,\mu)=\pi n_\mu \ .
  \end{equation}

\end{theo}
\begin{proof}
  The proof of equation \eqref{jost50} is standard and can be found in
  e.g.\ \onlinecite{Newton60,Newton86,Reed79III,Taylor83}. For sake of completeness we recall  
the main arguments. Integrate the logarithmic
  derivative of $F$ with respect to $k$ over a closed semi--circle in
  the complex upper half plane centered at the origin. According to the
  preceding lemma the integrand has simple poles on the positive
  imaginary axis each of which corresponds to a bound state of
  $h_\mu$. While the contribution of the semi--circle vanishes in the
  limit of infinite radius (see \eqref{pot3}) the integration over the
  real axis yields the phase difference by equation \eqref{jost48}.
  Applying the residue theorem
  completes the proof.
\end{proof}
We comment on the results of the above theorem. Equation \eqref{pot6}
generalizes Levinson's theorem to continuous angular momentum $\mu$.
One might wonder what happens when $\mu$ varies. On physical grounds
one expects that for higher angular momentum $\mu$ there are fewer
bound states since then the centrifugal barrier is stronger. Thus,
when $\mu$ is steadily increased the zeros of the Jost function $F$
move along the positive imaginary axis towards the origin. Every time
a zero moves out of the upper half plane we get a discontinuous jump
in the phase shift indicating the disappearance of a bound state.
\begin{theo}\label{theo:jost:3}
  Let $\mu\not\in\Z,\,\mu>0$ and assume $V$ satisfies \eqref{jost25}. In case
  $F(k=0,\mu)=0$ the above form of Levinson's theorem remains valid
  when $\mu>1$ and is modified as
  \begin{equation}
    \label{jost51}
    \delta(k=0,\mu)-\delta(k=\infty,\mu)=\pi (n_\mu+\mu) \ 
  \end{equation}
when $\mu<1$.
\end{theo}
So far we have not been able to settle the case $\mu\in\Z$
corresponding to boson and fermion statistics. For real anyons, however, note
that in case of positive angular momentum we have $\mu=\alpha$ when
$\mu<1$. Thus, equation \eqref{jost51} then specializes to
\begin{equation}\label{inverse}
\delta(k=0,\alpha)-\delta(k=\infty,\alpha)=\pi (n_\mu+\alpha),
\end{equation}
showing that one can determine the statistics parameter from the
scattering phase independently of the detailed form of the short range
potential $V$. Below we
will find a similar formula for the delta potential. Setting
$\alpha=1/2$ (i.e. for semions) in \eqref{jost51} we obtain
the well known resonance case of three dimensional Schr{\o}dinger
theory\cite{Newton60}, since then the radial Schr{\o}dinger equation
is formally identical to the one in three dimensions (compare
\eqref{jost2} and \eqref{jost3}).
\begin{proof}
  The strategy for proving \eqref{jost51} is similar to the one used for
  \eqref{jost50}. The techniques used are analogous to those in three 
  dimensional Schr\"odinger theory (see \onlinecite{Newton60,Taylor83}) but
  the line of argument is slightly changed. \\
  Because $F(k=0,\mu)=0$ we  change the integration contour as
  follows. We cut the semi-circle at the origin 
  and insert a small semi-circle $C_\rho$ of radius
  $\rho\ll 1$ centered at the origin. Then there is an additional
  contribution to equation 
  \eqref{jost50} namely,
  \begin{equation}\label{jost51a}
    \pi n'_\mu=\delta(k=0,\mu)-\delta(k=\infty,\mu)+
    \lim_{\rho\to 0}\frac{1}{2i}\oint_{C_\rho}d\ln F,
  \end{equation} 
  where $n'_\mu$ is the number of bound states with the exception of a
  possible zero energy eigenstate. In order to compute the second term
  of the right hand side in the above equation we only need to know the
  small energy behavior of $F$. Since we have assumed that the potential
  is decaying exponentially we can continue $F$ in the lower half plane and 
  expand it into a convergent power series around the origin for suitably small
  $k$. We start by rewriting expression \eqref{jost33},
  \begin{eqnarray}\label{jost52}
    F(k,\mu ) &=& 1+\tfrac{1}{k}\,
    \skal{\chi^{-}_{0}(\cdot\;;k,\mu )}{V\phi(\cdot\;;k,\mu)}\notag\\
    &=& 1+\tfrac{1}{k\sin\pi\mu}\bigl[
    \skal{\phi_0(\cdot\;;k,-\mu )}{V\phi(\cdot\;;k,\mu)}\bigr.\notag\\
    &&\bigl.\qquad\qquad\qquad-e^{-i\pi\mu}\skal{\phi_0(\cdot\;;k,\mu
      )}{V\phi(\cdot\;;k,\mu)}\bigr]. 
  \end{eqnarray}
  Here we have used the identity $H^{(1)}_\mu(z)=\tfrac{1}{i\sin\pi\mu}
  [J_{-\mu}(z)-e^{-i\pi\mu}J_\mu(z)]$ which can be found in
  e.g. \onlinecite{Watson22,Magnus66}. Note that in case of integer $\mu$ the Hankel function
  is defined by the limit of r.h.s. of the last equation. Then the above
  decomposition \eqref{jost52} is no longer valid, whence we restrict
  ourselves to the non-integer case. Both integrals in \eqref{jost52}
  converge uniformly as can be shown by use of the estimates 
  $$|\phi_0(r;k,\mu)\phi(r;k,\mu)|\leq C(\mu)e^{2|\Imt k|\,r}
    \left(\frac{|k|r}{1+|k|r}\right)^{2\mu+1}\qquad\qquad$$
  $$\qquad\qquad|\phi_0(r;k,-\mu)\phi(r;k,\mu)|\leq C(\mu)e^{2|\Imt k|\,r}
  \left(\frac{|k|r}{1+|k|r}\right),$$
  which follow from \eqref{jost15} and \eqref{jost16}. The functions
  $\phi_0(r;k,-\mu)$ and  $\phi_0(r;k,\mu)$ can be written as a
  convergent power series in $k^2$ times a factor
  $k^{-\mu+\frac{1}{2}}$ and $k^{\mu+\frac{1}{2}}$ respectively. This
  follows from  their definition \eqref{jost4} and the series
  expansions of the Bessel functions $J_{-\mu}(kr),J_\mu(kr)$, see
  e.g. \onlinecite{Watson22,Magnus66}. From \eqref{jost13} we see that also
  $\phi(r;k,\mu)$ can be written as a convergent power  series in
  $k^2$ times $k^{\mu+\frac{1}{2}}$, since the Green's function is an
  even function in $k$. Therefore, in a suitably small chosen neighborhood of 
  $k=0$ the Jost function can be expanded in a convergent series of the form
  \begin{equation}\label{jost53}
    F(k,\mu)=\sum_{n=0}^\infty a_n(\mu)k^{2n}+
    k^{2\mu}\sum_{n=0}^\infty b_n(\mu)k^{2n},\quad |k|\ll 1.
  \end{equation}
  Recall our convention to place te branch cut of $F$ along the
  negative imaginary axis. The assumption $F(0,\mu)=0$ gives $a_0=0$. In order to deduce from this 
  expansion the small energy behavior of $F$ we need to know the first 
  non-vanishing coefficient in \eqref{jost53}. We claim 
  $a_1(\mu)\neq 0$ for $\mu>1$ and $b_0(\mu)\neq 0$ for $\mu<1$. But then we have
  \begin {displaymath}\label{jost54}
    F(k,\mu)=\left\{\begin{array}{lcl}
        {\cal O}(k^{2\mu}) & \text{if} & \mu< 1\\
        {\cal O}(k^2) & \text{if} & \mu> 1
      \end{array}\right.. 
  \end {displaymath} 
  Thus, evaluating the integral over the semi-circle $C_\rho$ yields
  \begin{equation}\label{jost55}
    \lim_{\rho\to 0}\frac{1}{2 i}\oint_{C_\rho}d\ln F
    =\left\{\begin{array}{lcl}
        -\pi\mu & \text{if} & \mu<1\\
        -\pi & \text{if} & \mu> 1
      \end{array}\right.. 
  \end{equation}  
  In order to prove the claim we
  consider the limit $\lim_{k\to 0}k^{-1}\frac{\partial}{\partial
  k}F$. If we can show that
  \begin {displaymath}\label{jost56}
  \lim_{k\to 0}k^{-1}\frac{\partial}{\partial k}F(k,\mu)=\left\{
    \begin{array}{lcl}
      \infty & \text{if} & \mu<1\\
      C(\mu)\neq 0 & \text{if} & \mu> 1
    \end{array}\right. 
  \end {displaymath}
  then the claim follows as can be infered directly from \eqref{jost53}. We calculate the
  expression $\frac{\partial}{\partial k}F$ by making use of the
  identity \eqref{jost31}. Since the solutions $\chi^+,\phi$ are ill
  defined for $k\to 0$ we consider instead the modified solutions
  \begin {displaymath}\label{jost57}
  \Psi(r;k,\mu):=k^{\mu-\frac{1}{2}}\chi^+(r;k,\mu)\quad\text{and}\quad
  \Phi(r;k,\mu):=k^{-\mu-\frac{1}{2}}\phi(r;k,\mu)
  \end {displaymath}
  which are both finite and non-vanishing at $k=0$. This can be seen
  by taking the limit $k\to 0$ in the defining integral equations
  \eqref{jost10} and \eqref{jost21} respectively. In particular one then derives
  that $\Phi(r;0,\mu)$ behaves like $r^{\mu+\frac{1}{2}}$ for $r\to 0$
  and $\Psi(r;0,\mu)$ like $r^{-\mu+\frac{1}{2}}$ for
  $r\to\infty$. In terms of $\Phi,\Psi$ the Jost function and its 
  derivative w.r.t. $k$ (denoted by a dot) then read 
  \begin {displaymath}\label{jost58}
  F=W(\Psi,\Phi)\quad\text{and}\quad
  \dot F=W(\dot\Psi,\Phi)+W(\Psi,\dot\Phi).
  \end {displaymath}
  Since
  $\Psi$ is a solution of the radial equation one easily verifies the identity
  $$\frac{\partial}{\partial r}W[\Psi(r;k,\mu),\Psi(r;k',\mu)]=
  (k^2-{k'}^2)\Psi(r;k,\mu)\Psi(r;k',\mu).$$
  Differentiating w.r.t. $k$ and setting subsequently $k=k'$ one
  obtains from this the relation
  \begin{equation}\label{jost59}
    W[\dot\Psi(r;k,\mu),\Psi(r;k,\mu)]=-2k\int_r^\infty dr'\Psi(r';k,\mu)^2,
  \end{equation}
  where the r.h.s. is finite for $\Imt k>0$. The assumption
  $F(0,\mu)=0$ implies that the
  regular and the irregular solution are proportional 
  when $k=0$, i.e. $\Psi(r;0,\mu)=\kappa(\mu)\Phi(r;0,\mu)$
  holds. Note that the proportionality factor $\kappa(\mu)$ is nonzero
  and finite because both solutions are non-vanishing. Thus, in the limit $k\to 0$ we have
  $$ \dot F(0,\mu)=\kappa(\mu)^{-1}W(\dot\Psi(r;0,\mu),\Psi(r;0,\mu))+
  \kappa(\mu)W(\Phi(r;0,\mu),\dot\Phi(r;0,\mu)).$$
  However, $\dot F(0,\mu)$ does not depend on $r$ whence we can set $r=0$
  in the last equation. Since the regular solution $\Phi$ vanishes at
  $r=0$ we obtain from \eqref{jost59},
 \begin{eqnarray}\label{jost60}
  \lim_{k\to 0}k^{-1}\frac{\partial}{\partial k}F(k,\mu)&=&
  \lim_{k\to 0}k^{-1}\kappa(\mu)^{-1}W[\dot\Psi(0;k,\mu),\Psi(0;k,\mu)]\notag\\
  &=&-2\kappa(\mu)^{-1}\int_0^\infty dr\,\Psi(r;0,\mu)^2\notag\\
  &=&-2\kappa(\mu)\int_0^\infty dr\,\Phi(r;0,\mu)^2 
  \end{eqnarray}
  The irregular solution $\Psi$ falls off like
  $r^{-\mu+\frac{1}{2}}$ for large $r$ and is proportional to the
  regular solution $\Phi$ at $k=0$, therefore the r.h.s. of \eqref{jost60} is
  finite for $\mu>1$. This means that the regular solution is square
  integrable and hence a proper eigensolution of the full Hamiltonian
  $h_\mu$. The number of bound states is thus $n_\mu=n_\mu'+1$,
  whence we read of from \eqref{jost51a} and \eqref{jost55} that
  Levinson's theorem also holds in case $F(0,\mu)=0$ when $\mu>1$.\\ 
  For $\mu< 1$, however, the regular solution is no longer square
  integrable and the expression \eqref{jost60} is divergent. Hence the
  number of bound states is given by $n_\mu=n_\mu'$ and from
  \eqref{jost51a} and \eqref{jost55} 
  we infer the modified equation \eqref{jost51}. This completes the proof. 
\end{proof}

\section{The $\delta$--Potential}        \label{sec:delta}
In this section we will introduce and discuss the $\delta$--potential,
often also called the contact potential or zero--range interaction
potential because it describes an interaction which is
``non--vanishing'' only if the two particles are at the same place,
i.e. if $z_{rel}=0$. In the literature this figures also under the
notion of anyons without hard--core condition \cite{Kim98}. See e.g.\ 
\onlinecite{Manuel91,Bourdeau92,Amelino94} for previous articles on
this subject also in connection with field theoretic considerations.

To introduce the $\delta$--potential we follow a standard
strategy (see e.g.\ \onlinecite{Albeverio88}), i.e. we consider the
free Hamiltonian (in the center of mass system) restricted to a
definition domain of wave functions which vanish at $z_{rel}=0$. The
resulting operator has deficiency indices (1,1) and so there is
a one--parameter family of self--adjoint extensions, one of which is of
course the ordinary free Hamiltonian given as the Friedrich's
extension.  We discuss the bound state problem and the resulting
scattering theory. We also check the validity of Levinson's theorem. 
\Absatz
For a start, consider the operator
\begin{displaymath}
  h_{0,\mu} = -\partial_r^2 + \frac{\mu^2-4^{-1}}{r^2} 
\end{displaymath}
on $\cL^2(\R^+,dr)$ (recall~(\ref{jost1})) with domain of definition
${\cal D}(h_{0,\mu})$ consisting of wave functions $\psi (r)$ having
compact support away from the origin such that $\psi$ and $\partial_r
\psi$ are locally absolutely continuous and such that $h_{0,\mu} \psi$
(defined in the sense of distributions) is an element of
$\cL^2({\R}^+,dr)$.  Recall that locally absolutely continuous
functions are such that their derivatives in the sense of
distributions are locally integrable functions (see
e.g.~\onlinecite{Thirring79}).

Let us now investigate the most general solution of the equation
$h_{0,\mu}\psi = 0 $, given by
\begin{displaymath}
  c_1\cdot r^{\frac{1}{2} +\mu} + c_2\cdot r^{\frac{1}{2}-\mu} \ . 
\end{displaymath}
Using Weyl's terminology (consult e.g.\ \onlinecite{Weidmann76}), one has
the limit point case at $r=\infty$. At $r=0$ one also has the limit
point case if and only if $\mu\ge 1$, otherwise one has the limit
circle case.  If now $0\le \mu < 1$, by Weyl's alternative there is
exactly one solution of $h_{0,\mu}\psi=0$, which is $\cL^2$ near
infinity.  By construction this solution must be $\cL^2$ at both 
infinity and at zero, and thus is follows that $\dim (\ker
(h_{0,\mu}\pm i)) =1$ by Weyl's alternative.  Therefore the deficiency
indices are (1,1) and there is a one--parameter family of s.a.\ 
extensions (see e.g.~\onlinecite{Reed79II}).

In terms of the statistics parameter $\alpha$ we have the following
situation: If one chooses $\alpha=1$, then $\mu=\modulus{2m+\alpha}\ge
1$ for all $m\in {\bf Z}$ and all the $\left\{h_{0,\mu}\right\}$ are
essentially self--adjoint.  Now, if $\alpha\in[0,1)$ we have $\mu< 1$
if and only if $m=0$. In this case, just as in the three dimensional
case, the $\delta$--potential affects only one of the angular momentum
channels, namely the one associated with the smallest eigenvalue in
the sense of the modulus (in three dimensions called the s--channel).

Let $\alpha$ be in $[0,1)$. To construct the s.a.\ extensions of the
operators $h_{0,\alpha}$ with domain ${\cal D}(h_{0,\alpha})$ given
above, following \onlinecite{Bulla85,Albeverio88} we define the regular and
irregular solutions of $h_{0,\alpha}\psi=0$:
\begin{align*} 
  F_\alpha(r) &:= r^{\frac{1}{2}+\alpha} &&\text{(regular)}
  \\
  G_\alpha(r) &:= F_\alpha(r) \int_r^{r_0} dr'\, (F_\alpha(r'))^{-2} &&
  \text{(irregular)}
\end{align*}
where $r_0\in\R^+$ may be chosen arbitrary. The choice $r_0=1$ gives
\begin{displaymath} 
  G_\alpha (r) =
  \begin{cases}
    \frac{1}{2\alpha} (r^{\frac{1}{2}-\alpha} -r^{\frac{1}{2}+\alpha}) 
    & \alpha>0 \\
    -\sqrt r \ln r & \alpha=0
  \end{cases} 
  \ \ .
\end{displaymath}
                               
In order to explicitly construct the s.a.\ extensions of the
$\left\{h_{0,\alpha}\right\}$ we take recourse to the following
theorem, to be found in e.g. \onlinecite{Weidmann76}.
\begin{theo}
  For $\alpha<1$ the operator $h_{0,\alpha}$ with domain ${\cal D}
  (h_{0,\alpha})$ has a one--parameter family of s.a.\ extensions $\{
  h_{0,\alpha} (s)\}_{s\in {\bf R}}$ which are essentially s.a.\ on the
  domains
  \begin{subequations}
  \begin{displaymath}\label{delta1}
    {\cal D} (h_{0,\alpha} (s)) =\{ \psi \in {\cal D} (h_{0,\alpha}^\ast )
    \mid \,\lim_{r\downarrow 0} W(\psi_{\alpha,s},\psi) (r) = 0 \}.
  \end{displaymath}
  Here again $W$ is the Wronskian and
  \begin{displaymath} \label{delta10}
    \psi_{\alpha,s} := G_\alpha + s F_\alpha.
  \end{displaymath}
\end{subequations}
The choice $s=\infty$ gives the Friedrich's extension.
\end{theo} 
\Absatz %
In appendix~\ref{sec:res} we will prove the following
\begin{theo} 
  \label{theo:delta:2}
  The integral kernel of the resolvent of $h_{0,\alpha}(s)$ for $\alpha<1$
  is given as
  \begin{multline}\label{delta2}
    \frac{1}{h_{0,\alpha} (s)-k^2} (r,r') = \\ =\frac{i\pi}{2}
    \sqrt{rr'} \, J_\alpha (kr_<) \, H_\alpha^{(1)} (kr_>)
    - A(k,\alpha;s) \cdot \sqrt{rr^\prime} \,
    H_\alpha^{(1)} (kr)  \, H_\alpha^{(1)} (kr')  \ ,
  \end{multline}
  where $k$ satisfies $0<\arg (k)<\pi$.  $A(k,\alpha;s)$ is given as:
  \begin{equation}
    \label{delta3}
    A(k,\alpha;s) = \frac{\pi}{2} \left[
    \left(\frac{k}{2}\right)^{-2\alpha} (2s\alpha-1)
    \frac{1}{\sin\pi\alpha}\frac{\Gamma (1+\alpha)}{\Gamma (1-\alpha)}
    +\cot\pi\alpha 
    -i \right]^{-1}
  \end{equation}
\end{theo}
Note that $\lim_{s\to\infty} A(k,\alpha;s)=0$ such that with
\eqref{delta2} in the limit $s\to\infty$ we indeed recover the kernel of the
resolvent of the Friedrich's extension~\eqref{jost8}. Taking the
``bosonic limit'' $\alpha\rightarrow 0$ one arrives at the
expression:
\begin{displaymath} 
  A(k,0;s) = \frac{\pi^2}{4} \left[\ln \frac{k}{2i} 
    +\gamma +s\right]^{-1} \qquad \text{($\gamma$: Euler's constant)}
\end{displaymath}
Combined with \eqref{delta2} we have arrived at an expression for the
resolvent of the $\delta$--potential in the bosonic case, which agrees
with~\onlinecite{Albeverio88}.

\Absatz We turn to the discussion of the bound states. It can be shown
that the essential spectrum as a set is equal to $\R^+_0$ and that
$\R^+$ is purely absolutely continuous. We therefore have to look for
poles of the resolvent at energies $E_b=k_b^2<0$.  By~\eqref{delta2}
we have to search for the poles of $A(k,\alpha;s)$ as a function of
$k$ and by~\eqref{delta3} this gives the condition
\begin{equation} \label{delta4}
  k_b^{2\alpha} = 2^{2\alpha} e^{i\pi\alpha\,\bmod\,2\pi i} \frac{\Gamma
    (1+\alpha)} {\Gamma (1-\alpha)} (1-2\alpha s).
\end{equation}
The restriction $0<\arg(k)<\pi$ now implies $0<\arg
(k^{2\alpha})<2\pi\alpha$. Applying this condition to \eqref{delta4} leads to
three different cases:
\begin{enumerate}
\item[(i)]
  $1-2\alpha s>0$: Then \eqref{delta4} may be solved to give
  \begin{displaymath}
    k_b = k_b (\alpha,s) = 2i \left( \frac{\Gamma (1+\alpha)}{\Gamma(1-\alpha)} 
      (1-2\alpha s)\right) ^{\frac{1}{2\alpha}} \ ,
  \end{displaymath}
  i.e. $k_b$ is purely imaginary. The case $k_b=0$ occurs when $\alpha
  \uparrow 1$.  Note that this requires $0\le s<\frac{1}{2}$. Hence
  there is exactly one bound state.
  \label{delta9}
\item[(ii)]
  $1-2\alpha s<0$: Since by definition $0\le\alpha <1$, it is easy to see
  that in this case there is no solution to equation~\eqref{delta4},
  only in the limit $\alpha\uparrow 1$ one obtains $k_b=0$. This 
  means that there is no bound state. 
  
\item[(iii)]
  $1-2\alpha s=0$: This immediately gives $k_b=0$, which implies that
  there is no bound state.
\end{enumerate}

For comparison we recall that in the three dimensional case the
attractive $\delta$--potential supports at most one bound state. A
bound state exists if the particles obey bosonic statistics. For
fermions there is none. In the present context a bound state always exists for
$\alpha=0$ (bosons) and disappears at the latest when $\alpha=1$
(fermions). The fact that for $\alpha=0$ there is always a bound state
suggests that in two dimensions there is no such thing as a repulsive
$\delta$--potential. 

Below we argue, that case (iii) corresponds to a
zero energy resonance in three dimensional Schr{\"o}dinger theory.
Figure~\ref{bound state energies} gives a plot of $\modulus{k_b(\alpha,s)}$ as a
function of $s$ for various values of $\alpha$ in case (i), i.e. when
$1-2\alpha s>0$.
To calculate the bound--state wave function $\psi_b$ when $1-2\alpha
s<0$, we note that
\begin{displaymath}
  \frac{1}{h_{0,\alpha} (s) -k^2} \approx \frac{1}{k_b^2 -k^2} P_b
\end{displaymath}
for $k\approx k_b$ where $P_b$ is the orthogonal projector onto the
eigenspace of energy $E_b=k_b^2 <0$. Since this pole of the resolvent
is isolated, $P_b$ may be calculated as a residue. The calculation is
easy and shows that $P_b$ is indeed one--dimensional and that the
associated (normalized) eigenfunction $\psi_b$ is given as
\begin{displaymath}
  \psi_b(r;\alpha) = \left( \frac{2\sin\pi\alpha}{\pi\alpha} \right)^\frac{1}{2} 
  \sqrt{r\modulus{E_b}} K_\alpha (\modulus{k_b}r)
\end{displaymath}
for $\alpha<1$ and hence $\modulus{k_b(\alpha,s)}>0$.  For any fixed $\alpha<1$
the wave function decays exponentially for large $r$, which guarantees
square--integrability. In the limit $\alpha\uparrow 1$ the bound
state wave function approaches $r^{-\frac{1}{2}}$ times a normalization
constant, which is not square integrable. The case $\alpha=1$ should
therefore be considered as a resonance at $E=0$.

We turn to a discussion of the resulting scattering theory.  First we
note that due to relation~\eqref{delta2} the difference of the
resolvents for $h_{0,\alpha}(s)$ and the free Hamiltonian
$h_{0,\alpha}(s=\infty)$ is a rank--one operator and therefore in
particular trace class. Therefore the $S$--matrix exists and is
unitary by the Kuroda--Birman theorem (see e.g.\ \onlinecite{Kuroda62}).
\begin{theo}
  For given $\alpha$ and $E=k^2>0$ the outgoing states for the
  Hamiltonian $h_{0,\alpha}(s)$ are given as
  \begin{displaymath} 
    \phi^{+}(r;k,\alpha)=\phi_{0}(r;k,\alpha)
    +\frac{2i}{\pi} A(k,\alpha;s) \cdot \chi_{0}^{+}(r;k,\alpha)\ .
  \end{displaymath}
  In particular the resulting partial wave amplitude and the phase shift
  take the form
  \begin{subequations}
    \begin{eqnarray}
      \label{delta5}  f(k,\alpha;s) &=& \frac{4i}{\pi}
      \frac{A(k,\alpha;s)}{\sqrt{\pi i k}}\ , \\
      \label{delta6} e^{2i\delta (k,\alpha;s)} &=& 1 + \frac{4i}{\pi}
      A(k,\alpha;s)\ .
    \end{eqnarray}
  \end{subequations}
\end{theo}
\begin{proof}
  We start by rewriting the resolvent \eqref{delta2} as
  \begin{equation} 
    \label{delta7}
    \frac{1}{h_{0,\alpha}(s)-k^{2}}(r,r') =\left(\phi_{0}(r_{<};k,\alpha)+\frac{2i}{\pi}
    A(k,\alpha;s)\chi_{0}^{+}(r_{<};k,\alpha)\right) \frac{i}{k}
    \chi_{0}^{+}(r_{>};k,\alpha) \ .
  \end{equation}
  Now we observe that (see e.g.~\onlinecite[p.37]{Albeverio88})
  \begin{equation} 
    \label{delta8}
    \lim_{\varepsilon \downarrow 0}\lim_{r'\rightarrow \infty} 
    e^{-i(k+i\varepsilon)r'}
    \frac{1}{h_{0,\alpha}(s)-(k+i\varepsilon)^{2}}(r,r')
  \end{equation}
  must be proportional to the outgoing solution $\phi^{+}(r;k,\alpha)$. The
  claim now follows by applying \eqref{delta8} to \eqref{delta7}.
\end{proof}
\Absatz 
In view of Levinson's theorem we would like to make the following
comments on this last result: The limiting values of
$\delta(k,\alpha;s)$ at $k\rightarrow 0^+$ and $k\rightarrow\infty$
are \em a priori \em determined up to an additive constant in
$\pi\cdot\Z$, only. We shall choose these constants in such a way,
that $\delta(k,\alpha;s)$ as a function of $k\in\R^+$ can be defined
as a continuous function of $k$.  Now, $\exp(i2\delta(k,\alpha;s))$ is
of the form $(a(k,\alpha;s)+i)(a(k,\alpha;s)-i)^{-1}$. Here
$a(k,\alpha;s)$ is a real--valued, continuous function of $k$.
Explicitly
\begin{displaymath}
  a(k,\alpha;s) = \frac{\pi}{2} A(k,\alpha;s)^{-1} + i
\end{displaymath}
(This observation also shows that the
right hand side of \eqref{delta6} is indeed unimodular.) Hence
\begin{displaymath}
  \delta(k,\alpha;s) = \arg(a(k,\alpha;s)+i) \ .
\end{displaymath}
Here the branch of $\arg(\cdot)$ is chosen such that
$\delta(k,\alpha;s)$ is a continuous function of all its arguments. In
particular we can achieve $\delta(k,\alpha;s)\in(0,\pi)$, for any $k$
and fixed $\alpha<1$ and $s$. For the cases, where there is a bound
state, this is consistent with the attractive nature of the
interaction.  By the explicit form of $a(k,\alpha;s)$ it is easily
seen, that
\begin{align*}
  \delta(k,\alpha;s) \quad &\xrightarrow{k\rightarrow 0^+} \quad 
  \pi\frac{1-\sgn(2\alpha s-1)}{2} \\
  \delta(k,\alpha;s) \quad &\xrightarrow{k\rightarrow \infty\,} \quad \pi\alpha 
\end{align*}
The fraction $n:= \frac{1-\sgn(2\alpha s-1)}{2}$ takes values in
$\{0,1\}$ and is in fact nothing but the number of bound states given
by~\eqref{delta4}. We therefore have the following relation
\begin{equation}\label{inverse-delta}
  \delta(0^+,\alpha;s) - \delta(\infty,\alpha;s) = \pi(n-\alpha) \ ,
\end{equation}
which is \em not \em of the form one would naively expect in view
Levinson's theorem~\ref{theo:jost:2}.  In this context we note, that
the statement of Levinson's theorem also fails in the non--anyonic
three dimensional case, if a $\delta$--interaction is
considered. However, we obtain a formula similar to \eqref{inverse}
relating the phase shift to the statistics parameter $\alpha$.

\Absatz By relations \eqref{cross15} and \eqref{jost41a}, we obtain
the following expression for the differential cross--section:
\begin{displaymath}
  \frac{d\sigma}{d\theta}=
  \modulus{f(k,\alpha;s) \cdot e^{-i\pi\alpha} +
  f_{AB}(k,\theta)}^2 \ ,
\end{displaymath}
where $f_{AB}$ and $f$ are given by expression~\eqref{qm24} and
\eqref{delta5}, respectively. Inserting these explicit expressions,
one obtains ---away from the forward direction--- the following
relation 
\begin{displaymath}
  \frac{d\sigma}{d\theta} = \frac{1}{\pi k}
  \modulus{\sin\pi\alpha\, \left(\cot\theta-i\right) +
  e^{-i\pi\alpha} \frac{4i}{\pi} A(k,\alpha;s)}^2 \ .
\end{displaymath}
So far, we have been unable to analyze this in a non--numerical
way. Figures~\ref{cross--sections for contact interaction 1} and
\ref{cross--sections for contact interaction 2} show
some numerical results of the differential cross--section as a function of the
scattering angle $\theta$. Displayed is the deviation from pure
Aharonov--Bohm scattering which is asymptotically reached when
$s\to\pm\infty$. Note the simultaneous cross over at
$\theta=\pi\alpha$ for varying $s$ in
each of the plots with $s\neq\pm\infty$. In particular, the point 
corresponding to the cross over always lies on the graph
($s=\pm\infty$) associated with the Aharonov--Bohm cross--section. We 
like to point out that in both figures~\ref{cross--sections for
  contact interaction 1} and \ref{cross--sections for contact interaction 2} only for $s=0,2$ and $\alpha=0.5$ (semions) the
differential cross--section actually vanishes for some scattering
angle $\theta$. In the other cases the
cross--section becomes very small but is non--vanishing for all $\theta$. 

\section{The Square Well Potential}        \label{sec:step}
Another explicitly solvable problem in scattering theory is given by the 
square well,
\begin{displaymath}
V(r)=\left\{\begin{array}{ll}
                -V_0, &  r\leq d\\
                0, &  r>d
                \end{array}\right.\quad\text{with}\quad V_0>0.
\end{displaymath}
The regular solution of the eigenvalue problem $(h_\mu-k^2)\psi=0$
in the interval $0<r\leq d$ is given by
$$\phi(r;k,\mu)=\left(\tfrac{k}{q}\right)^{\mu+\frac{1}{2}}\phi_0(r;q,\mu),
\quad\text{where}\quad q=\sqrt{k^2+V_0}.$$ 
For $r>d$ the eigensolutions of $h_\mu$ obey the free radial 
equation and hence the Jost solutions in this interval are just given by the 
free solutions,
$$\chi^\pm(r;k,\mu)=\chi_0^\pm(r;k,\mu),\quad r>d.$$
Making the usual requirement that all solutions are continuously 
differentiable we can evaluate the Wronskian $W(\chi^\pm,\phi)$ at $r=d$ and 
obtain the following expression for the Jost function,
\begin{equation*}
F(k,\mu)=\left(\tfrac{k}{q}\right)^{\mu-\frac{1}{2}}
\left[\chi_0^+(d;k,\mu)\phi_0(d;q,\mu-1)-\tfrac{k}{q}\chi^+_0(d;k,\mu-1)
\phi_0(d;q,\mu)\right].
\end{equation*}
Here, we have used the common differentiation rules for Bessel and Hankel 
functions (see e.g.\ \onlinecite{Watson22,Magnus66}). With the help of the Jost 
function we can now calculate the phase shift and the partial wave
amplitude. We find 
\begin{displaymath}
\tan \delta(k,\mu)=i\,\frac{F-\bar F}{F+\bar F}=
\frac{q J_\mu(kd)J_{\mu-1}(qd)-k J_{\mu-1}(kd)J_\mu(qd)}{
q Y_\mu(kd)J_{\mu-1}(qd)-k Y_{\mu-1}(kd)J_\mu(qd)}
\end{displaymath} 
\begin{displaymath}
f(k,\mu)=\frac{\bar F-F}{\sqrt{\pi ik}F}=\frac{-2}{\sqrt{\pi ik}}
\,\frac{q J_\mu(kd)J_{\mu-1}(qd)-k J_{\mu-1}(kd)J_\mu(qd)}{
qH^{(1)}_\mu(kd)J_{\mu-1}(pd)-kH^{(1)}_{\mu-1}(kd)J_\mu(qd)}.
\end{displaymath}
Inserting the last expression into the formula \eqref{jost41a} for the
partial wave decomposition of the scattering amplitude we can
calculate the differential cross sections for various parameters
$\alpha$. Note that for small momenta we can make use of the small energy
behavior of the partial wave amplitude which is given by
\begin{displaymath}
k\to 0:\qquad f(k,\mu)={\cal O}(k^{2\mu-1}).
\end{displaymath}  
Thus, when calculating the scattering amplitude in the partial wave 
decomposition the partial wave amplitudes for higher angular momentum hardly 
contribute to the infinite sum in \eqref{jost41a}. According to our discussion in section \ref{sec:cross} we have to add the Aharonov--Bohm scattering amplitude $f_{AB}$ in order to obtain the differential cross--section (see 
\eqref{cross15}). A numerical example is displayed in figure
\ref{square well cross--sections}. 
For different values of the 
statistics parameter $\alpha$ one obtains very different angular 
distributions of the scattered particles. One observes the interpolation between the bosonic ($\alpha=0$) and the 
fermionic ($\alpha=1$) cross section at energy $E=k^2=0.25$. Note that the singularities at the 
scattering angles $\theta=0$ and $\theta=\pi$ for $\alpha\not\in\Z$ enter due to  
the long range nature of the statistics. In fact, according to 
\eqref{cross15} they result from adding the Aharonov-Bohm amplitude given 
in \eqref{qm24}. For  fermions ($\alpha=1$) there is no scattering under an 
angle of $\theta=\pi/2$ as is well known. The figure for $\alpha=0.5$ shows 
that a similar effect can take place for fractional statistics.
\Absatz
According to our discussion in 
section \ref{sec:Jost} the zeros of the Jost function $F$ lie on the positive 
imaginary axis and determine the bound states. Thus, plotting the solutions of 
\begin{displaymath}
  F(ik,\mu)=0,\quad k>0
\end{displaymath}
we get so called (real) {\em Regge trajectories} (see e.g. \onlinecite{Newton86}) 
displaying the $\mu$ dependence of the discrete
spectrum of $h_\mu$. Figure \ref{Regge trajectories for the square well} provides a numerical example. In
contrast to three dimensional scattering theory here every point on
the Regge trajectories has a specific physical meaning because the
statistics parameter $\alpha$ interpolates between the integer valued
angular momentum channels.

\section{Conclusion}                    \label{sec:conc}

We have demonstrated that it is possible to formulate nonrelativistic
quantum scattering theories for two particles obeying anyon statistics. In
particular we have proven the existence of the wave operators under certain
assumptions concerning the interaction. For spherically symmetric
potentials we gave criteria for completeness, i.e.\ unitarity of the
$S$--matrix. These criteria turned out to be the same as in three
dimensional Schr{\"o}dinger theory. It remains an open problem to show
completeness in the general case in a way comparable to Enss' method.
To generalize the latter, we believe it is crucial to obtain a better
analytic control of the propagator. (There is an extensive literature on
a similar problem, namely to obtain a closed form of the Aharonov--Bohm
propagator see e.g. \onlinecite{Gerbert89}.) 
\Absatz
We extended the notion of differential cross--section to two
dimensions and showed that in case of anyon statistics, the
corresponding scattering amplitude consists of two parts. The first 
encodes the anyon statistics and resembles the Aharonov--Bohm
amplitude, while the second is relevant to spectral analysis of the
perturbed Hamiltonian. For spherically symmetric potentials we carried
out this analysis by introducing Jost functions and showed how
fractional statistics is equivalent to fractional angular momentum. We
also showed that Levinson's theorem holds in the conventional case 
and gave the modifications necessary in the presence of a zero energy
resonance. In the latter case we found that for positive angular
momentum $\mu<1$ the statistics parameter -- independently of the
detailed form of the short range potential -- can be determined from the scattering
phase, namely (compare \eqref{inverse})
$$\alpha\equiv\frac{1}{\pi}\Bigl(\delta(k=0,\alpha)-\delta(k=\infty,\alpha)\Bigr)\mod\Z.$$
We then applied our results to the following two examples. First we
considered the $\delta$--potential, where we extracted information
regarding the point spectrum of the perturbed Hamiltonian from its
resolvent and showed that the above relation between statistics
parameter and scattering phase is also valid in this context(compare
\eqref{inverse-delta}). Second we considered the square well potential, where we
obtained similar information from the corresponding Jost function. In
both situations we numerically evaluated the differential
cross--section for non--integer values of the statistics parameter.
It then became evident that fractional statistics is fundamentally
different from Bose and Fermi statistics: The statistical gauge forces
produce singularities of the differential cross--section in forward direction,
showing the long range nature of these forces. Due to this fact the
angular dependence of the statistical interaction will invariably show
up in the scattering data. In particular this might cause a
differential cross--section which is constant for bosons or fermions,
but which becomes angular dependent for intermediate statistics. Our
discussion of the $\delta$--potential provides an example.
\Absatz
The results presented in this paper may also be relevant to the
investigation of bulk properties of anyon matter, via the relation
between the virial coefficients and the scattering data \cite{Beth37}.
This we intend to discuss in a forthcoming publication.

\acknowledgements
The authors would like to thank Jens Mund for
numerous discussions concerning the differential geometric formulation
of fractional statistics and valuable comments on the results of this
paper. This work was supported in part by DFG Sfb 288 ''Differentialgeometrie 
und Quantenphysik''.

\newpage
\appendix

\section{The Bundle Theoretic Formulation}
\label{sec:bun}
In section~\ref{sec:qm} we used the physical picture of particles
carrying flux--tubes in order to describe the anyon model. Anyons,
however, are but a special case in a more general class of particles,
called plektons, obeying neither Bose nor Fermi statistics. In this
appendix we give a short review of plektonic systems, i.e. systems of
$n$ particles whose statistics is determined by a finite dimensional
unitary representation of the braid group $B_n$. The appropriate
mathematical tool to describe such particles is the concept of vector
bundles. This will in particular provide the Hilbert space and a
canonical ``free'' dynamics described by a certain Laplace operator. Our presentation will closely follow the one given in
\onlinecite{Mund92,Mund95}, where, however, emphasis was mainly given to
relativistic formulations using momentum space considerations. For
other references on the bundle theoretic formulation see e.g. \onlinecite{Horvathy89,Imbo90}.  We
continue to use $\C$ to describe the one--particle configuration
space.  Let $\C^{\times n}$ denote the $n$--fold product, viewed as
the configuration space for $n$ distinguishable particles. The set
$D_n$ in $\C^{\times n}$ is the set of all points
$z=(z_1,z_2,\ldots,z_n)$ with $z_i=z_j$ for at least one pair of
different indices. Any element $\sigma$ of the permutation group $S_n$
acts in an obvious way on $\C^{\times n}$ via
$(z\sigma)_i=z_{\sigma(i)}$. We define the configuration space of $n$
identical particles in two dimensions to be
\begin{displaymath}
  \nC :=(\C^{\times n}\setminus D_n)/S_n \ ,
\end{displaymath}
with points in it denoted by $\underline z$. Let $\NC$ denote the
universal covering space of $\nC$ with points there being written as
$\tilde{z}$ and be $\pi : {\NC} \rightarrow {\nC}, \tilde{z} \mapsto
\pi (\tilde{z}) = \underline{z}$ the associated projection mapping.

It is well--known that the fundamental group of $\nC$, denoted by
$\pi_1(\nC)$, is isomorphic to the braid group $B_n$ and that any
element $b\in B_n$ acts in a standard way from the right $\tilde
z\mapsto \tilde z b$ on the manifold $\NC$.  Furthermore, the
universal covering $\NC\mapsto\nC$ can be viewed as a principal fiber
bundle over $\nC$ with structure group $\pi_1(\nC)$ and fiber
$\pi^{-1} (\underline z)$ for any $\underline z\in\nC$. 
Given any finite dimensional unitary representation
$b\mapsto\rho(b)$ of $B_n$ in a Hilbert space $F$ with scalar product 
$\skal{\cdot}{\cdot}$ there is an associated hermitian vector bundle. This vector bundle ${\cal F}$ with
base space $\nC$ is given by $\NC\times_{\rho,B_n}F$, which by
definition is the set of orbits in $\NC\times F$ under the following
action of $B_n$ on this space
\begin{equation}
  \label{bun2}
  b:(\tilde z ,f)\longmapsto\left(\tilde z b, \rho(b^{-1}) f\right)\ ,
  \qquad\tilde z \in\NC,\ f\in F.
\end{equation}
${\cal F}$ is a smooth fibered space with base $\nC$ and fibers
isomorphic to $F$. On the fiber over $\underline{z}$ there is a natural
scalar product denoted by $\skal{\cdot}{\cdot}_{\underline{z}}$.
Furthermore there is a canonical measure $d\mu (\underline{z})$ on
$\nC$ inherited from the Lebesgue measure on $\C^{\times n}$.  This
defines a scalar product on the space $\Gamma_{c} ({\cal F})$ of
smooth sections in ${\cal F}$ with compact support via
\begin{equation}
  \label{bun3}   
  \skal{\psi}{\phi} := \int\limits_{\nC}
  \skal{\psi(\underline{z})}{\phi(\underline{z})}_{\underline{z}}
  d\mu(\underline{z}) \ .
\end{equation}
By $\cL^2({\cal F})$ we denote the resulting Hilbert space completion.
There is another Hilbert space equally well suited and canonically
isomorphic to this space. Consider the set of maps $ \Psi:{\NC}\mapsto
F$ which are smooth with $\pi(\supp\Psi)\subset{\nC}$ being compact
and which satisfy the equivariance property
\begin{equation}
  \label{bun4}
  \Psi(\tilde{z}b)=\rho(b^{-1})\Psi(\tilde{z}) ,\qquad 
  \forall \ \tilde{z}\in{\NC},\ b \in B_n \ .
\end{equation}
For any two such functions $\Phi$ and $\Psi$ we therefore have
\begin{equation}
  \label{bun5}      
  \skal{\Psi(\tilde{z}b)}{\Phi(\tilde{z}b)} =
  \skal{\Psi(\tilde{z})}{\Phi(\tilde{z})} \qquad \forall\ \tilde
  z\in\NC,\ b\in B_n \ .
\end{equation}
Hence this expression depends on $\underline{z}=\pi(\tilde{z})$ only
and so the integral over the base space makes sense and we may define
a scalar product by
\begin{equation}
  \label{bun6}
  \skal{\Psi}{\Phi} :=\int\limits_{\nC}
  \skal{\Psi(\tilde{z})}{\Phi(\tilde{z})} d\mu(\underline{z}).
\end{equation}
The resulting Hilbert space obtained again by completion is denoted by
$\cL^2_{eq}(\NC,F)$ and there is a canonical isomorphism between
$\cL^2({\cal F})$ and $\cL^2_{eq}(\NC,F)$ (see e.g.\ 
\onlinecite{Mund92,Mund95}). Furthermore the canonical (flat) Levi--Civita
connection on $\C^{\times n}$ induces a (flat) connection on $\nC$
which in turn defines a hermitian connection $\nabla$ on ${\cal F}$.
The associated Bochner or generalized Laplacean (see e.g.\ \onlinecite{Berline92})
$\Delta= \nabla\circ\nabla$ on $\cL^2({\cal F})$ is what defines a
free Hamiltonian $H_0= -\Delta/2m$ for a system of $n$ plektons, if
$m>0$ is taken to be the mass of one particle.
\Absatz %
When the unitary representation $\rho$ of $B_n$ is one--dimensional
one speaks of anyons. Any such representation
is obviously abelian and can be shown to be of the form
$b_k\mapsto\exp(i\alpha\pi)$ in terms of the generators $b_1, \dots
b_{n-1}\in B_n$ for a fixed $\alpha\in[0,2)$. This follows easily from
the observation that all the $b_k$'s are conjugate to each other. We
denote the resulting line bundle by $\cF_\alpha$.                                 
Another consequence of $F$ being one--dimensional is that all anyonic
line bundles are actually trivial. This observation was first made by
J.~S.~Dowker~\cite{Dowker85} based on V.~I.~Arnold's result that
$H^2(\nC,\Z)=0$ \cite{Arnold70} and the classification theorem of
Cartan, Kostant, Souriau and Isham. It was rediscovered by
M.~Gaberdiel~\cite{Gaberdiel} and is implicitly contained in \onlinecite{Wu84} and
\onlinecite{Froehlich91}. Another proof is given in \onlinecite{Mund92,Mund95}.                               \Absatz 
Let us now consider in more detail the Hilbert spaces constructed
above when $n=2$. We first introduce relative coordinates by
considering the following transformation of $\C^{\times 2}$:
\begin{equation}
  \label{bun9}
  (z_1,z_2)\longmapsto \left( 2^{-1}(z_1+z_1),z_1-z_2\right) =
  (z_{cen},z_{rel}) 
\end{equation}
The transposition in $S_2\cong\Z_2$ obviously maps $(z_{cen},z_{rel})$
into $(z_{cen},-z_{rel})$. Therefore $^2\C$ is diffeomorphic to
$\C\times\cone$. The first factor is the configuration space for the
center of mass motion. The corresponding quantum mechanical discussion
is analogous to ordinary multi particle systems, since it is not
affected by the statistics. Therefore we will concentrate on the
second factor and the associated quantum mechanical description.                                
In the same way as $\R$ is the universal covering space of $S^1$, the
universal covering space of $\cone\cong\R^+\times S^1$ is given by
$\R^+\times\R$ with the projection mapping
\begin{equation}
  \label{bun10}
  \pi :\R^+\times\R \longrightarrow \R^{+}\times S^{1} \quad \text{,
  with} \qquad (r,\theta)\longmapsto (r,e^{2i\theta}) \ .
\end{equation}
The action of $B_{2}$ on the universal covering is given in terms of
its only generator $b_1$ as
\begin{equation}
  \label{bun11}
        (r,\theta)\longmapsto (r,\theta)b_1 := (r,\theta +\pi)\ .
\end{equation}
Choosing the representation $b_1\mapsto\exp(-i\pi\alpha)$ for a fixed
$\alpha\in [0,1]$ we are prepared to
construct the line bundle $\cal F_\alpha$ and the Hilbert space
$\cL^2(\cal F_\alpha)$ or equivalently $\cL^2_{eq}(\NC,F)$. The fact
that the bundle $\cal F_\alpha$ is trivial is now reflected in the
possibility to ``pull--down'' the theory from either of those $\cL^2$
spaces onto the space of square integrable functions on $\nC$ itself,
denoted by $\cL^2(\nC)$. Explicitly there is the following unitary map
from $\cL^2(\nC)$ onto $\cL_{eq}^2(\NC)$ given as
\begin{displaymath}
  \psi\in\cL^2(\nC)\longmapsto \Psi\in\cL^2_{eq}(\NC) \qquad
  \text{with} \qquad \Psi(\tilde z) = e^{i\pi\alpha\cdot \theta(\tilde
  z)} \cdot \psi\left(\pi(\tilde z)\right) \ ,
\end{displaymath}
where $\theta(\tilde z)$ is a real valued, continuous function of
$\tilde z$ induced by the polar angle. With the help of this unitary mapping, we can pull--down
the Bochner Laplacean from $\cL^2(\cal F_\alpha)$ to obtain an
expression for the free Hamiltonian on $\cL^2(\nC)$. The result of
this procedure is precisely the Hamiltonian~\eqref{qm1} we introduced
in section~\ref{sec:qm} on physical grounds.  

Let it be noted that $\alpha$ was introduced as a parameter fixing the
representation $B_2\ni b_1\mapsto\exp(-i\pi\alpha)$ and is hence
determined up to an additive even integer, only.  Therefore we loose
no generality if we restrict $\alpha$ to the interval $[0,2)$.  In
fact, we have $H_0(\alpha+2) = \exp(-i2\theta) H_0(\alpha)
\exp(i2\theta)$, which is a gauge transformation reflecting the
arbitrariness we have. On the other hand, time inversion causes
$\alpha$ to change sign, i.e.\ if $T$ denotes the anti--unitary
operator of time inversion, we have $H_0(-\alpha)=T H_0(\alpha) T$.
However $H_0(-\alpha)$ is in turn equivalent to $H_0(2-\alpha)$ and
consequently the anyon models with $\alpha\in[0,1]$ are connected to
those with $\alpha\in[1,2]$ by time reversal. This justifies the
restriction of $\alpha$ to the interval $[0,1]$.

Furthermore we remark that $\cL^2(\cone,rdrd\theta)$ is unitary
equivalent to the square integrable \em symmetric \em functions on the
punctured plane $P_+\cL^2(\C^\star,\frac{1}{2}rdrd\theta)$ and
therefore the case $\alpha=0$ corresponds to the bosonic case (since
$H_0(\alpha=0)=-\triangle$). If $\alpha=1$ our model is equivalent
to $H_0(\alpha=1)$ acting on $P_+ \cL^2(\C^\ast)$, which is in turn
unitary equivalent to $-\triangle$ on $P_-
\cL^2(\C^\ast,\frac{1}{2}rdrd\theta)$, the Hilbert space of square
integrable, \em anti--symmetric \em functions on the punctured
plane. Hence we call anyons obeying statistics corresponding to
$\alpha=1$ fermions.

\Absatz %
There is also a geometric description of the universal covering space
essentially as the Riemann surface of the function $\log (z)$, which
is useful for the understanding of the Hilbert space constructions
made above and which goes as follows.  Take an infinite set of upper
and lower half planes in $\C$ and glue them together alternatively
along the positive and negative axes respectively.  In fact, consider
an element $f$ in $C^{\infty}_{0,\alpha}(\HH\setminus\{0\})$.  Then it
is easy to see that this $f$ uniquely defines a function $\Psi$ in
$C^{\infty}_{0,eq}(\NC,F_{\alpha})$ and conversely any function $\Psi$
in $C^{\infty}_{0,eq}(\NC,F_{\alpha})$ defines a unique element $f$ in
$C^{\infty}_{0,\alpha} (\HH\setminus\{0\})$. Furthermore this
correspondence is linear and the scalar products are the same under
this correspondence. This establishes the desired isomorphism between
${\cal H}_{\alpha}$ and $\cL^2_{eq}(\NC,F_{\alpha})$ and hence also
$\cL^2({\cal F}_{\alpha})$. This isomorphism also extends to the
various Laplace operators. Indeed, note first that locally $\NC$ looks
like $\C^{\times n}$, so there is a natural Laplace operator
$\tilde{\Delta}$ on $C^{\infty}(\NC)$, the space of smooth functions
on $\NC$. This operator is easily seen to map
$C^{\infty}_{0,eq}(\NC,F_{\alpha})$ into itself and hence defines an
operator on $\cL^2_{eq}(\NC,F_{\alpha})$.  Furthermore this operator
also corresponds to the Laplace operator on
$C^{\infty}_{0}(\HH\setminus\{0\})$ under the above correspondence
$f\leftrightarrow\Psi$. Finally it is easy to see that under the
canonical isomorphism between $\cL^2_{eq}(\NC,F_{\alpha})$ and
$\cL^2({\cal F}_{\alpha})$ referred to above $\tilde{\Delta}$
corresponds to the Bochner Laplacean $\Delta=\nabla\circ\nabla$.
\Absatz

\section{Some Estimates} \label{sec:est}
This appendix is devoted to a proof of the lemmas~\ref{lem:pot:2} and
\ref{lem:pot:3}.  
\begin{proof}[Proof of lemma~\ref{lem:pot:2}:]
  We start with a proof of lemma~\ref{lem:pot:2}.  For the
  moment let $0<\epsilon<1$ be arbitrary. Since $\modulus{e^{ix}-1} \leq
  \min(2,\mid x\mid)$, we have $\modulus{e^{ix}-1} < 2 \modulus
  x^{\epsilon}$ for all $ x\in \R.$ This gives
  \begin{align*}
    \modulus{ I_{\alpha}(\rho ,\chi )-I_{\alpha}(0,\chi)} &\le \modulus{
      \int_{-\infty}^{+\infty}dy\,\left(e^{i\rho\cosh y}-1 \right)
      \frac{e^{-\alpha y}}{1-e^{-2y-2i\chi}}} \\ \nonumber &\le 2
    \rho^{\epsilon}\int^{\infty}_{0}dy\,\cosh^\epsilon y \modulus{
      \frac{e^{-\alpha y}}{1-e^{-2y-2i\chi}}+ \frac{e^{\alpha
          y}}{1-e^{2y-2i\chi}}} \ .
  \end{align*}
  To proceed further,we split the integral into a part from 0 to 1 and
  the rest. This gives
  \begin{displaymath}
    \modulus{I_{\alpha}(\rho ,\chi )-I_{\alpha}(0,\chi)}\le
    2\rho^{\epsilon} \left( \cosh(1)\cdot I_{1}(\chi)\,+I_{2}(\chi )
    \right) \ .
  \end{displaymath}
  For $I_1$ we have an expression involving the sum of three integrals
  $I_{1,1}$, $I_{1,2}$ and $I_{1,3}$, given by
  \begin{multline*}
    I_1(\chi )= I_{1,1}(\chi)+I_{1,2}(\chi) + I_{1,3}(\chi) = \\ =
    \int_0^1 dy\,\modulus{\frac{e^{-\alpha y}-1}{1-e^{-2y-2i\chi}}} +
    \int_{0}^{1}dy\, \modulus{\frac{e^{\alpha y}-1}{1-e^{2y-2i\chi}}} +
    \\ + \int_0^1 dy\,\modulus{
      \frac{1}{1-e^{-2y-2i\chi}}+\frac{1}{1-e^{2y-2i\chi}}} \ .
  \end{multline*}
  For $I_2$ we have the expression
  \begin{displaymath}
    I_2(\chi) = \int_1^\infty dy\, \cosh^\epsilon y
    \left(\modulus{\frac{e^{-\alpha y}}{1-e^{-2y-2i\chi}}} +
      \modulus{\frac{e^{\alpha y}}{1-e^{2y-2i\chi}}} \right) \ .
  \end{displaymath}
  The aim is to estimate these quantities for $\chi\in [-\pi ,+\pi ]$.
  To estimate $I_{1,1}(\chi )$ we use the following little estimates:
  \begin{align}  
    \label{est1}
    \modulus{ 1-e^{-\alpha y}} &\le 2y && 
    \text{for $y>0$ and $\alpha\in(0,1)$:}
    \\
    \label{est2}
    \modulus{1-e^{-2y-2i\chi}} &\ge \modulus{ \,1-e^{-2y}} && 
    \text{for $y\in\R$:} 
  \end{align} 
  This gives
  \begin{displaymath}
    I_{1,1}(\chi) \le \int^1_0 dy\, \frac{\alpha y}{1-e^{-2y}} < C \ .
  \end{displaymath}
  $I_{1,2}(\chi )$ is estimated similarly, if one replaces the estimate
  \eqref{est1} by
  \begin{displaymath}
    0 \le e^{\alpha y}-1 \le 2 e^{2} y \ , \qquad 
    \text{ for } 0 \le y \le 1 \text{ and } \alpha\in (0,1) \ .
  \end{displaymath}
  To estimate $I_{1,3}(\chi )$ for $\chi\in [-\pi ,+\pi ]$ we first
  observe that for $\min(\modulus \chi, \pi -\chi,\pi +\chi) \ge
  \frac{\pi}{4}$ it is obviously bounded.  Hence it suffices to estimate
  the three remaining cases $\modulus \chi \le \frac{\pi}{4}$,
  $\frac{3}{4}\pi \le \chi \le \pi$ or $-\pi \le \chi \le -\frac{3}{4}
  \pi$. We only consider the first case, since the other two cases may
  be discussed with similar arguments.  So let $\modulus\chi \le
  \frac{1}{4}\pi$ and add and subtract $(\pm 2y+2i\chi)^{-1}$ in the
  integrand of $I_{1,3}(\chi)$. This way we can estimate $I_{1,3}$ by a
  sum of three integrals, which we will denote by $\hat I_i$.
  \begin{align*}
    I_{1,3}(\chi ) &\le \hat I_1(\chi )+\hat I_2(\chi
    )+\hat I_3(\chi ) = & \\ \nonumber &=
    \int^1_0 dy\,\modulus{\frac{1}{1-e^{-2y-2i\chi}}
      -\frac{1}{2y+2i\chi}} &+\int^1_0 dy\,\modulus{
      \frac{1}{1-e^{2y-2i\chi}} -\frac{1}{-2y+2i\chi}} \\ &
    &+\int^1_0 dy\,\modulus{\frac{1}{2y+2i\chi}+
      \frac{1}{-2y+2i\chi}}
  \end{align*}
  To estimate $\hat I_1(\chi)$ and $\hat I_2(\chi)$ we note that the
  function
  \begin{displaymath}
    G(\zeta)=\frac{1}{1-e^{-\zeta}}-\frac{1}{\zeta}
  \end{displaymath}
  is obviously analytic in $\left\{ \zeta\in\C \mid\modulus{\Im(\zeta)}
    \le \frac{\pi}{2}\right\}$ except possibly at the origin $\zeta =0$.
  However, $G(\zeta =0)=1$ and hence by Riemann's theorem, $G(\zeta)$ is
  an analytic function in that domain. In particular $G(\zeta)$ is
  bounded on every compact subset and this gives the boundedness of
  $\hat{I}_{1}(\chi )$ and $\hat{I}_{2}(\chi )$ for $\chi\in
  [-\frac{1}{4}\,\pi ,\frac{1}{4}\,\pi]$.  $\hat{I}_{3}(\chi )$ can be
  estimated as follows
  \begin{displaymath}
    \hat I_3(\chi)\le\int^1_0 dy\,
    \frac{\modulus\chi}{\chi^{2}+y^{2}}
    \le\,\int^\infty_0 d\eta\,\frac{1}{1+\eta^{2}}<\infty \ .
  \end{displaymath}
  This concludes the estimate for $I_{1}(\chi )$ and it remains to
  estimate $I_{2}(\chi )$. By \eqref{est2} we have
  \begin{displaymath}
    I_{2}(\chi )\,\le\,\int^{\infty}_{1}dy\,2\,e^{\epsilon y}
    \left(\frac{e^{-\alpha y}}{1-e^{-2y}}\,+\frac{e^{\alpha
          y}}{e^{2y}-1}\right) \ .
  \end{displaymath}
  Now this integral is finite whenever $\epsilon < \min (\alpha
  ,1-\alpha)$.
\end{proof}
\Absatz 
\begin{proof}[Proof of lemma~\ref{lem:pot:3}:]
  We will consider the third partial derivatives w.r.t.\ E only, since
  this is actually the case we will need and since the other cases may
  be discussed similarly. By interchanging integration and
  differentiation we formally have
  \begin{equation}
    \label{est3}
    \frac{\partial^{n}}{\partial E^{n}}\,v^{m,m'}_{\alpha}(E,E')
    =\int\limits_{\HH\,\setminus\{0\}}
    \frac{\partial^{n}}{\partial E^{n}}\left[\overline{\phi_{\alpha ;m,E}(z)}
      \,V(z)^{2}\,\phi_{\alpha ;m',E'}(z)\right] d\mu (z)
  \end{equation}
  for $n=1,2,3$. This is permitted provided the integrand is a
  measurable function in $z$, $E$ and $E'$, bounded in the sense of the
  modulus by a function which is in $\cL^1$ w.r.t\ $z$ and with uniform
  bounds in $E$ and $E'$ in compact sets in $(0,\infty)$. To prove this
  claim it obviously suffices to replace $\frac{\partial}{\partial E}$
  by $\frac{\partial}{\partial k}$ with $E=k^2$ in \eqref{est3}. By the
  explicit form of $\phi_{\alpha ;m,E}$ given by~\eqref{qm8}, for fixed
  $\alpha$, $m$ and $m'$ we therefore have to estimate the product
  \begin{displaymath}
    \modulus{\partial^{n}_{k}\,J_{\mu}(kr)}\cdot\modulus{ 
      J_{\mu}(k'r)}\cdot\modulus{V(r,\theta)}^2 \ , 
    \qquad \text{for $n=1,2,3$} \ .
  \end{displaymath}
  Using the formula to be found in \cite{Watson22} or \cite{Magnus66}
  \begin{equation}
    \label{est4}
    \frac{d}{dk}J_{\sigma}(kr)\,=\,\frac{r}{2}\,
    \left( J_{\sigma -1}(kr)-J_{\sigma +1}(kr)\,\right)
  \end{equation}
  it follows by iteration that it suffices to estimate
  \begin{equation}
    \label{est5}
    r^n \modulus{J_{\mu +l}(kr)} \cdot \modulus{J_{\mu '}(k'r)} \cdot 
    \modulus{V(r,\theta )}^2
  \end{equation}        
  for $-n\le l\le n$ with $0\le n\le 3$. Let the compact set in question
  be given by $E_{1}\le E,E'\le E_{2}$ such that $k_{1}\le k$, $k'\le
  k_{2}$.  Using the well known estimates
  \begin{align*}
    a) && \modulus{J_{\sigma}(kr)} &\le C(\sigma)
    && \text{ for } \modulus{kr}\le 1
    \\
    b) && \modulus{J_{\sigma}(kr)} &\le C(\sigma)(kr)^{-\frac{1}{2}}
    && \text{ for } \modulus{kr} \ge 1
  \end{align*}
  we see that \eqref{est5} with $k_1 \le k$, $k' \le k_2$ is bounded by
  \begin{displaymath}
    C(\mu,\mu')\cdot r^n \modulus{V(r,\theta )}^2 \le C(\mu,\mu')\cdot 
    \max (1,k^{-1}_2)^3\modulus{V(r,\theta)}^2
  \end{displaymath}
  for $r\le k^{-1}_2$ and by
  \begin{displaymath}
    C(\mu,\mu')\cdot r^{n-1} k^{-1}_1\modulus{V(r,\theta )}^2
    \le C(\mu,\mu')\cdot \max (k_1^2,k^{-1}_1)\cdot r^2 \modulus{V(r,\theta )}^2
  \end{displaymath}
  for $r\ge k^{-1}_1$. Obviously~\eqref{est5} is bounded for $r$ in the
  interval $(k^{-1}_2,k^{-1}_1)$ uniformly for $k_1 \le k,k'\le
  k_2$. This proves the claim and by our previous remark this
  concludes the proof.
\end{proof}

\section{Bounds for Jost functions} \label{sec:est2}
In this section we will prove the estimates \eqref{jost16},\eqref{jost28},
\eqref{jost15} and \eqref{jost27} for the functions $\phi$ and $\chi^{\pm}$
and their power series coefficients $\phi^{n}$ and $\chi^{\pm,n}$.
Also we will show that the analyticity properties for $\phi$ and
$\chi^{\pm}$ also extend to their derivatives with respect to $r$.  We
start with two preparations. Observe first that \eqref{jost16} and
\eqref{jost28} hold when $V=0$, i.e. for $\phi=\phi_{0}$ and
$\chi^{\pm}=\chi^{\pm}_{0}$ respectively. Indeed for $\phi_{0}$ \eqref{jost16}
follows from the relations (see e.g.\ \cite{Watson22,Magnus66})
\begin{displaymath}
  J_{\mu}(z)\,\xrightarrow[z\to 0]{}\,C(\mu)z^{\mu}\quad\text{and}\quad
  J_{\mu}(z)\,\xrightarrow[\mid z\mid \to\infty]{}\,
  \frac{\cos (z-\frac{\pi}{2}(2\mu+1))}{\sqrt{\,\frac{\pi}{2}z}}.
\end{displaymath}
For $\chi^{\pm}_{0}$ the equation \eqref{jost28} follows from the 
following asymptotic behavior
\begin{displaymath}
H^{\pm}_{\mu}(z)\,\xrightarrow[z\to 0]{}\,C(\mu)z^{\mu}\quad\text{and}\quad
H^{\pm}_{\mu}(z)\,\xrightarrow[z\to\infty]{}\,
\frac{e^{\pm i(z-\frac{\pi}{2}\mu-\frac{1}{4})}}{\sqrt{\,\frac{\pi}{2}z}}.
\end{displaymath}
Secondly for $g_{0}$ (see \eqref{jost10}) one establishes analogously 
\begin{eqnarray}\label{estb1}
  0\leq r'\leq r\,:\quad |g_0(r,r';k,\mu )|\leq
  C(\mu )\,e^{|\Im(k)|(r-r')}\left(\tfrac{r}{1+|k|r}\right)^{\mu +\frac{1}{2}}
  \left(\tfrac{r'}{1+|k|r'}\right)^{-\mu+\frac{1}{2}},\\
\notag\\
  0\leq r\leq r'\,:\quad |g_0(r,r';k,\mu)|\leq
  C(\mu )\,e^{|\Im(k)|r-\Im(kr')}
  \left(\tfrac{r}{1+|k|r}\right)^{\mu +\frac{1}{2}}
  \left(\tfrac{r'}{1+|k|r'}\right)^{-\mu+\frac{1}{2}}.
\end{eqnarray}
Now \eqref{jost15} is proved by induction on $n$ as follows. Define $A_n$ by
\begin{displaymath}
  |\phi_n(r;k,\mu)|\,=\,e^{|\Im(k)|r}\,
  \left(\tfrac{r}{1+|k|r}\right)^{\mu+\frac{1}{2}}\,A_{n}(r;k,\mu )
\end{displaymath}
and it suffices to show that
\begin{displaymath}
  A_{n}(r;k,\mu)\leq|k|^{\mu+\frac{1}{2}}\,
   \frac{C(\mu)^{n+1}}{n!}\left[\int^r_0dr'\,
   \frac{r'|V(r')|}{1+|k|r'}\right]^n,
\end{displaymath}
where $C(\mu)$ is the maximum of the $C(\mu)$ in \eqref{jost16} and the
one in \eqref{estb1}. By the preparatory remarks \eqref{jost15} holds for
$n=0$, i.e. for $\phi_{0}=\phi_{n=0}$. To perform the induction step, by
construction we have
\begin{displaymath}
 \phi_n(r;k,\mu )=\int^r_0dr'\,g_0(r,r';k,\mu)V(r')\phi_{n-1}(r';k,\mu),
\end{displaymath}
which gives the inequality
\begin{displaymath}
  A_n(r;k,\mu)\leq C(\mu)\int^r_0dr'\,
  \frac{r'|V(r')|}{1+|k|r'}\,A_{n-1}(r';k,\mu),
\end{displaymath}
from which the induction step follows easily. This completes the proof
of \eqref{jost15} and \eqref{jost16} follows from it by summation.\\
The proof of the bounds \eqref{jost27} and \eqref{jost28} is similar and
we will consider the case $\chi^{-}$ only. Now write
\begin{displaymath}
  |\chi^-_n(r;k,\mu)|=\,e^{\Im(kr)}
  \left(\tfrac{r}{1+|k|r}\right)^{-\mu+\frac{1}{2}}B_{n}(r;k,\mu ),
\end{displaymath}
such that it suffices to prove
\begin{displaymath}
  B_{n}(r;k,\mu)\leq\frac{C(\mu )^{n+1}}{n!}\,
  |k|^{-\mu+\frac{1}{2}}
  \left(\int^{\infty}_{r}dr'\,\frac{r'|V(r')|}{1+|k|r'}\right)^{n}.
\end{displaymath}
By the preparatory remarks, this inequality is valid for $n=0$. To
make the induction step, we note that by construction
\begin{displaymath}
  \chi^{-}_n(r;k,\mu)=\int^{\infty}_{r}dr'\,g_{0}(r,r';k,\mu )\,
  \chi^-_{n-1}(r';k,\mu ).
\end{displaymath}
Hence one has the estimate
\begin{displaymath}
  B_{n}(r;k,\mu)\leq C(\mu )\,\int^{\infty}_{r}dr'\;
 \frac{r'|V(r')|}{1+|k|r'}\;B_{n-1}(r';k,\mu),
\end{displaymath}
from which the induction step easily follows. This completes the proof
of \eqref{jost27} and \eqref{jost28} follows by summation.

\section{The Resolvent of the $\delta$--Potential} \label{sec:res}
\begin{proof}[Proof of theorem~\ref{theo:delta:2}:]
  For any $\eta \in \cL^2(\R^+)$ define
  \begin{displaymath}
    \psi := \frac{1}{h_{0,\alpha}(s)-k^2}\, \eta \ ,
  \end{displaymath}
  using the integral kernel given by~\eqref{delta2}. It is easily seen
  that $\psi$ is well--defined. In order to prove the theorem we shall
  show that \eqref{delta2} formally defines a resolvent, i.e. we have 
  \begin{equation}
    \label{res1}
    \left( -\frac{\partial^2}{\partial
        r}-\frac{\alpha^2-\frac{1}{4}}{r^2} \right) \psi -k^2 \psi =
    \eta \ .
  \end{equation}
  That this ``resolvent property'' is formally satisfied, can easily
  be verified with the aid of the well--known formulae for the first
  derivatives of Bessel functions.  In particular one uses the
  relations:
  \begin{displaymath}
    C'_\alpha = \frac{\alpha}{z} C_\alpha - C_{\alpha +1} \qquad
    C'_{\alpha+1} = C_\alpha -\frac{\alpha+1}{z} C_{\alpha+1} \qquad
    \text{for}\quad C_\alpha = J_\alpha\text{ or } H_\alpha.
  \end{displaymath}
  Furthermore we shall show that $\psi$ satisfies the boundary
  condition in \eqref{delta1}, i.e.
  \begin{equation}
    \label{res2}
    \lim_{r\downarrow 0} W(\psi_{\alpha,s},\psi)(r) =0 \ .
  \end{equation}
  For convenience we introduce the following notation:
  \begin{align*} 
    I_<(r) &:= \int_0^r dr'\, \sqrt{r'}J_{\alpha}(kr')\eta (r') && I
    := \int_0^\infty dr'\, \sqrt{r'}H^{(1)}_{\alpha}(kr')\eta (r')
    \\
    I_>(r) &:= \int_r^\infty dr'\, \sqrt{r'}H^{(1)}_{\alpha}(kr')\eta
    (r') &&
  \end{align*}
  $\psi$ can now be cast into the form
  \begin{equation}
    \label{res3}
    \psi(r) = \sqrt r \left[ \frac{i\pi}{2} H^{(1)}_\alpha (kr)I_<(r)
      + \frac{i\pi}{2} J_\alpha(kr)I_>(r) -A(k,\alpha;s)
      H^{(1)}(kr) I\right] \ .
  \end{equation}
  To verify the boundary condition~\eqref{res2} we note that
  \begin{align*}
    && \psi_{\alpha,s}(r) &= \frac{1}{2\alpha}\cdot
    r^{\frac{1}{2}-\alpha} + \tilde s \cdot r^{\frac{1}{2}+\alpha}
    \qquad \text{with}\qquad \tilde s = s-\frac{1}{2\alpha}
    \\
    \text{giving} && \psi'_{\alpha,s}(r) &=\left(
      \frac{1}{4\alpha}-\frac{1}{2}\right)\cdot
    r^{-\frac{1}{2}-\alpha} +\left( \frac{\tilde s}{2}+\tilde s \alpha
    \right) r^{-\frac{1}{2}+\alpha} \\ \text{and hence} &&
    W(\psi_{\alpha,s},\psi)(r) &=B'(r) \left( \frac{1}{2\alpha}\cdot
      r^{1-\alpha} + \tilde s r^{1+\alpha}\right) + B(r) \left(
      \frac{1}{2}\cdot r^{-\alpha}-\alpha\tilde s r^\alpha\right) \ ,
  \end{align*}
  where $B(r)$ denotes the quantity in the square brackets in
  \eqref{res3}. In order to take the limit, we remark that
  $z^{-\alpha}J_\alpha(z)$ is an analytic function for any value of
  $\alpha\in\R$. From its power series expansion at $z=0$
  (see e.g.\ \cite{Magnus66}) one obtains the following asymptotic
  relations when $\alpha>0$:
  \begin{align*}
    \lim_{r\downarrow 0} r^{-\alpha} J_\alpha (kr) &= \iota_\alpha (k)
    := \left( \frac{k}{2} \right)^\alpha \frac{1}{\Gamma (1+\alpha)}
    \\
    \lim_{r\downarrow 0} \ \ r^\alpha Y_\alpha (kr) &= \gamma_\alpha
    (k) := - \left( \frac{2}{k} \right)^\alpha \frac{1}{\sin\pi\alpha}
    \frac{1}{\Gamma(1-\alpha)}
    \\
    \lim_{r\downarrow 0} r^\alpha Y_{-\alpha} (kr) &=
    \gamma_{-\alpha}(k) := -\left(\frac{k}{2}\right)^{-\alpha}
    \frac{1}{\tan\pi\alpha} \frac{1}{\Gamma(1-\alpha)}
  \end{align*}
  For the definition of the integral kernel~\eqref{delta2} we
  restricted the Bessel functions to the sheet given by
  $0<\arg(k)<\pi$.
  Since the irrational powers of $k$ in the above expressions stem
  from the asymptotic behavior of the Bessel functions, they have to
  be evaluated on the same sheet.  Bearing this in mind, we are
  prepared to take the limit $r\rightarrow 0^+$, which will make most
  of the terms in \eqref{res3} disappear. Since $\modulus{I_<(r)} \sim
  C\cdot r^{\alpha+1}$ and $I_> \sim I$ for $r\ll 1$, the surviving terms
  are:
  \begin{multline*}
    \lim_{r\downarrow 0} W(\psi_{\alpha,s},\psi)(r)= I \cdot
    \lim_{r\downarrow 0} \left[ \frac{i\pi}{2} r^{-\alpha} J_\alpha
      (kr) + \tilde s\cdot k\cdot A(k,\alpha;s) r^{1+\alpha}
      H^{(1)}_{\alpha+1}(kr) \right. \\ \left.  +\frac{k}{2\alpha}
      A(k,\alpha;s) r^{1-\alpha}
      \left(H^{(1)}_{\alpha+1}(kr)-\frac{2\alpha}{kr}
        H^{(1)}_{\alpha}(kr) \right) \right]
  \end{multline*}  
  Now, by a standard theorem for Bessel functions, we have
  $H^{(1)}_{\alpha+1}(kr)-\frac{2\alpha}{kr} H^{(1)}_{\alpha}(kr) = -
  H^{(1)}_{\alpha-1}(kr)$. The boundary condition~\eqref{res2}
  therefore implies
  \begin{displaymath}
    A(k,\alpha;s) = \frac{i\pi}{2} \iota_{\alpha}(k) \left[
      \iota_\alpha(k) -2i\tilde s \alpha \cdot \gamma_\alpha +
      \frac{ik}{2\alpha} \gamma_{\alpha-1} \right]^{-1} \ .
  \end{displaymath}
  Substituting $\tilde s = s-\frac{1}{2\alpha}$ and the expressions for $\iota_\alpha$, $\gamma_\alpha$ and
  $\gamma_{\alpha-1}$ we see that $A(k,\alpha;s)$ is given by
\eqref{delta3}, thus completing the proof of
theorem~\ref{theo:delta:2}.
\end{proof}


\bibliographystyle{amsplain}
                                
\bibliography{pub}
\newpage
\listoffigures
\newpage
\bildeinbindencm{14cm}{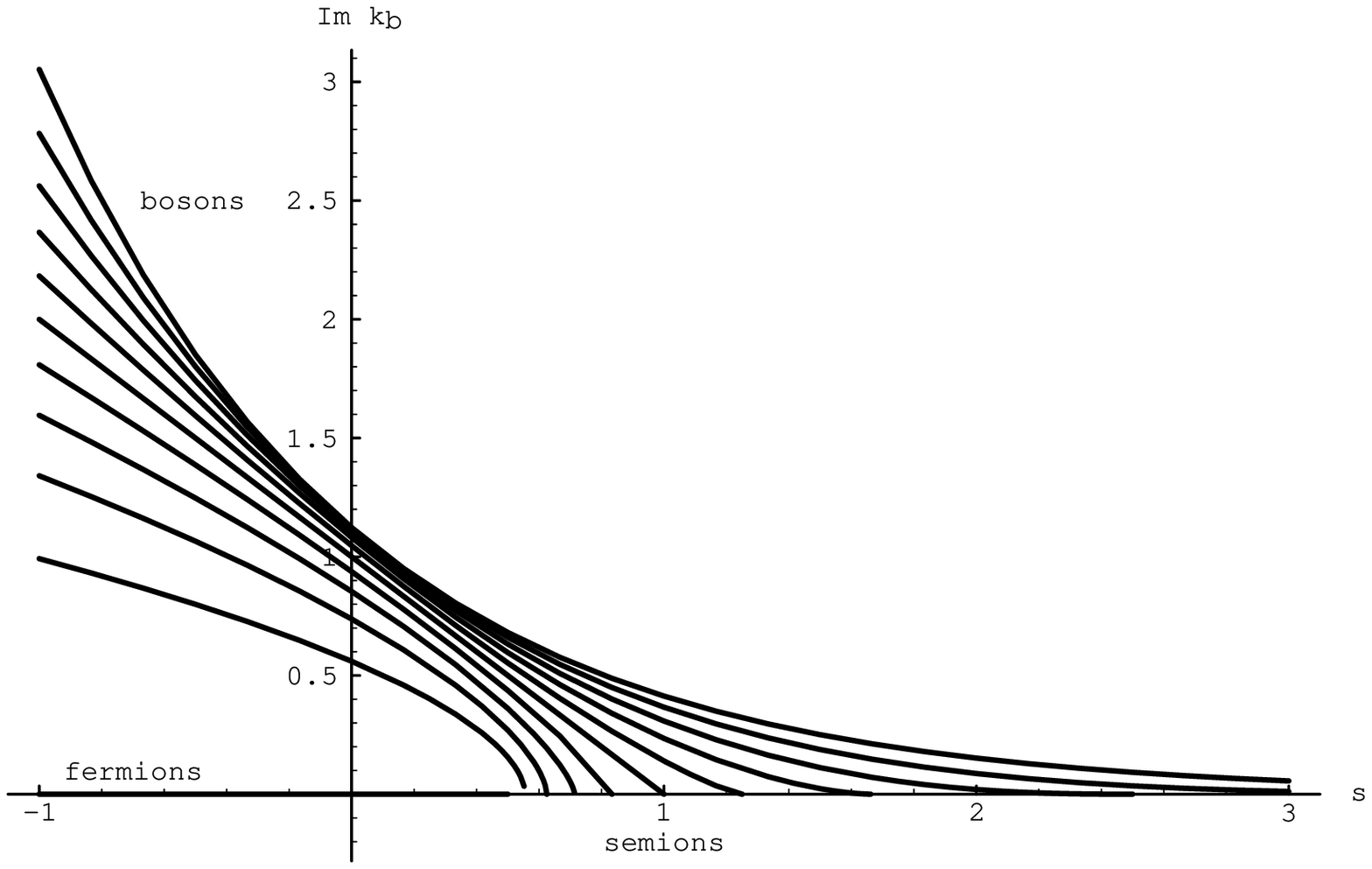}{bound state energies}{ The diagram shows the location of the square root of the bound state energy
  $k_b$ on the imaginary axis as a
  function of the coupling parameter $s$ for ten different, equispaced
  values of the statistical parameter $\alpha$.  ``bosons'' labels the
  graph corresponding to $\alpha=0$, ``semions'' corresponds to
  $\alpha=\tfrac{1}{2}$, the graph intersecting the $s$--axis at $1$
  and ``fermions'' corresponds to $\alpha=1$.  The latter case does
  not correspond to a bound state.  }
\newpage
\bildeinbindencm{13cm}{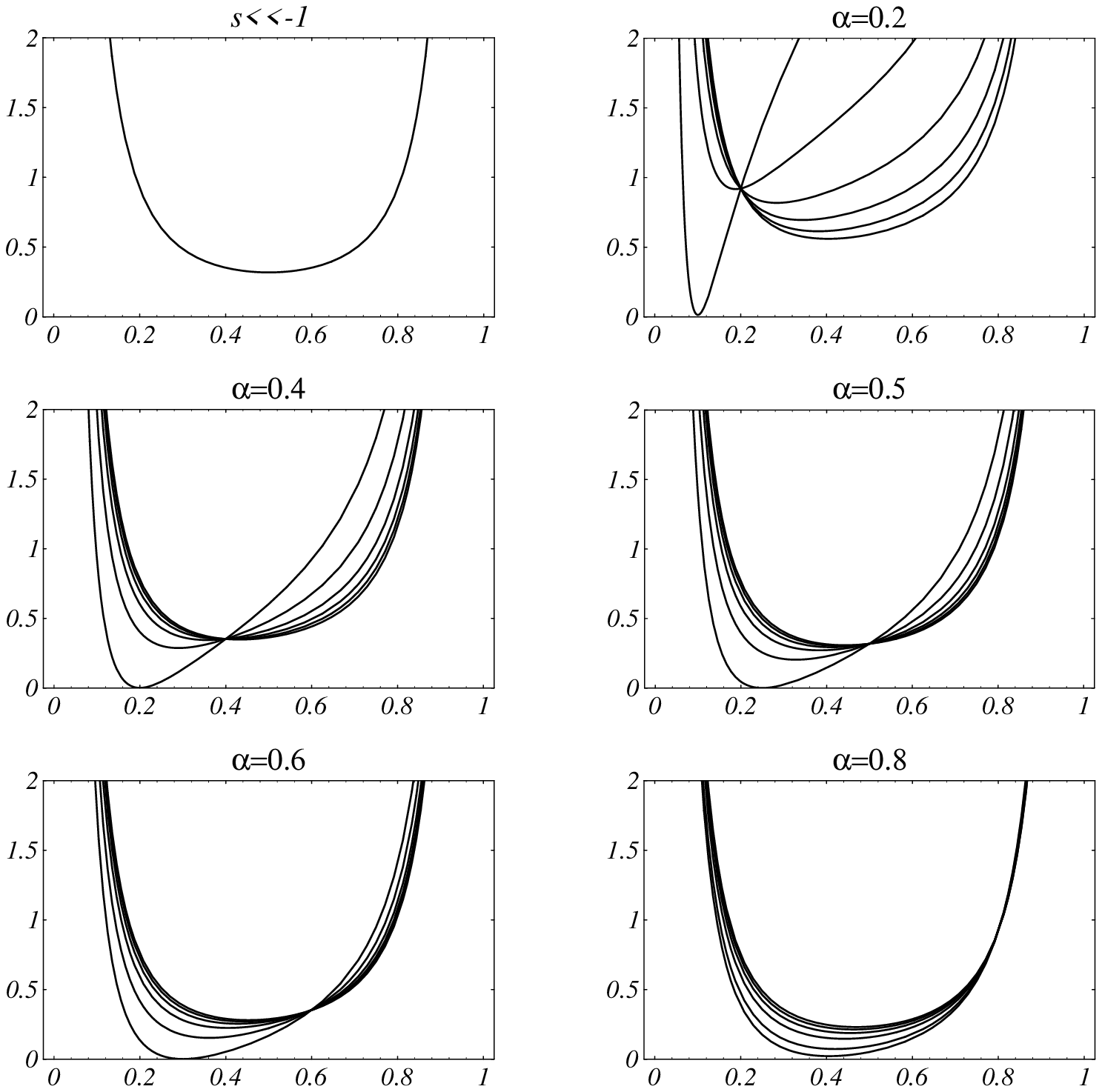}{cross--sections for contact
  interaction 1}{The
  above figures show the ``normalized'' differential cross--section
  $\tfrac{1}{\sin^2\pi\alpha}\tfrac{d\sigma}{d\theta}$ as a function 
  of the scattering angle $\theta$ in units of $\pi$. The
  normalization factor is chosen such that the asymptote in the limit
  $s\to-\infty$ (i.e. when the interaction is turned off, giving the
  Friedrich's extension of $h_{0,\alpha}$) becomes independent of
  $\alpha$. The latter is shown in the first plot (top left) 
  corresponds to Aharonov--Bohm scattering. The other five plots show
  a family of graphs corresponding to the following coupling strengths
  $s=-10,-8,\ldots,0$, $k=1$ and the given values for $\alpha$. For
  small values of $\theta$ the graphs for different $s$ are ordered
  from left to right starting with $s=0$.}
\newpage
\bildeinbindencm{13cm}{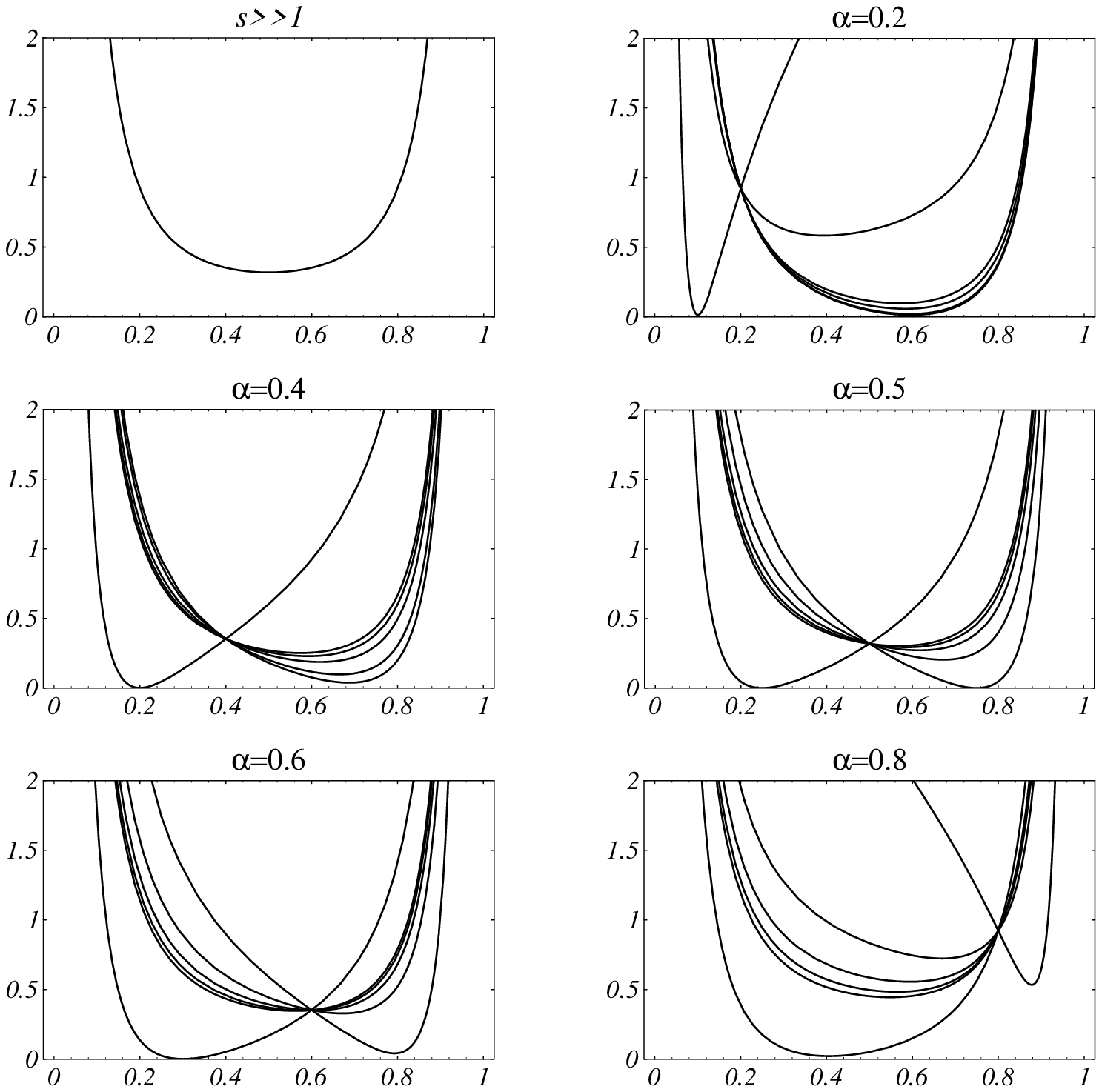}{cross--sections for contact
  interaction 2}{Similar to
  figure \ref{cross--sections for contact interaction 1} the above plots show the differential
  cross--section
  $\tfrac{1}{\sin^2\pi\alpha}\tfrac{d\sigma}{d\theta}$ as a function 
  of the scattering angle $\theta$ in units of $\pi$. But now only
  positive $s$ values are considered. The first plot (top left) shows
  the limiting case $s\to\infty$ corresponding again to 
  Aharonov--Bohm scattering. The other five plots display 
  a family of graphs with coupling strengths
  $s=0,2,\ldots,10$, $k=1$ and the given values for $\alpha$. Here the
  ordering of graphs is different from figure \ref{cross--sections for
  contact interaction 1}. For small values of $\theta$ the leftmost
  graph always corresponds to $s=0$ but the ordering is reversed once at
  a certain value $s=s(\alpha)$.}
\newpage
\bildeinbindencm{14cm}{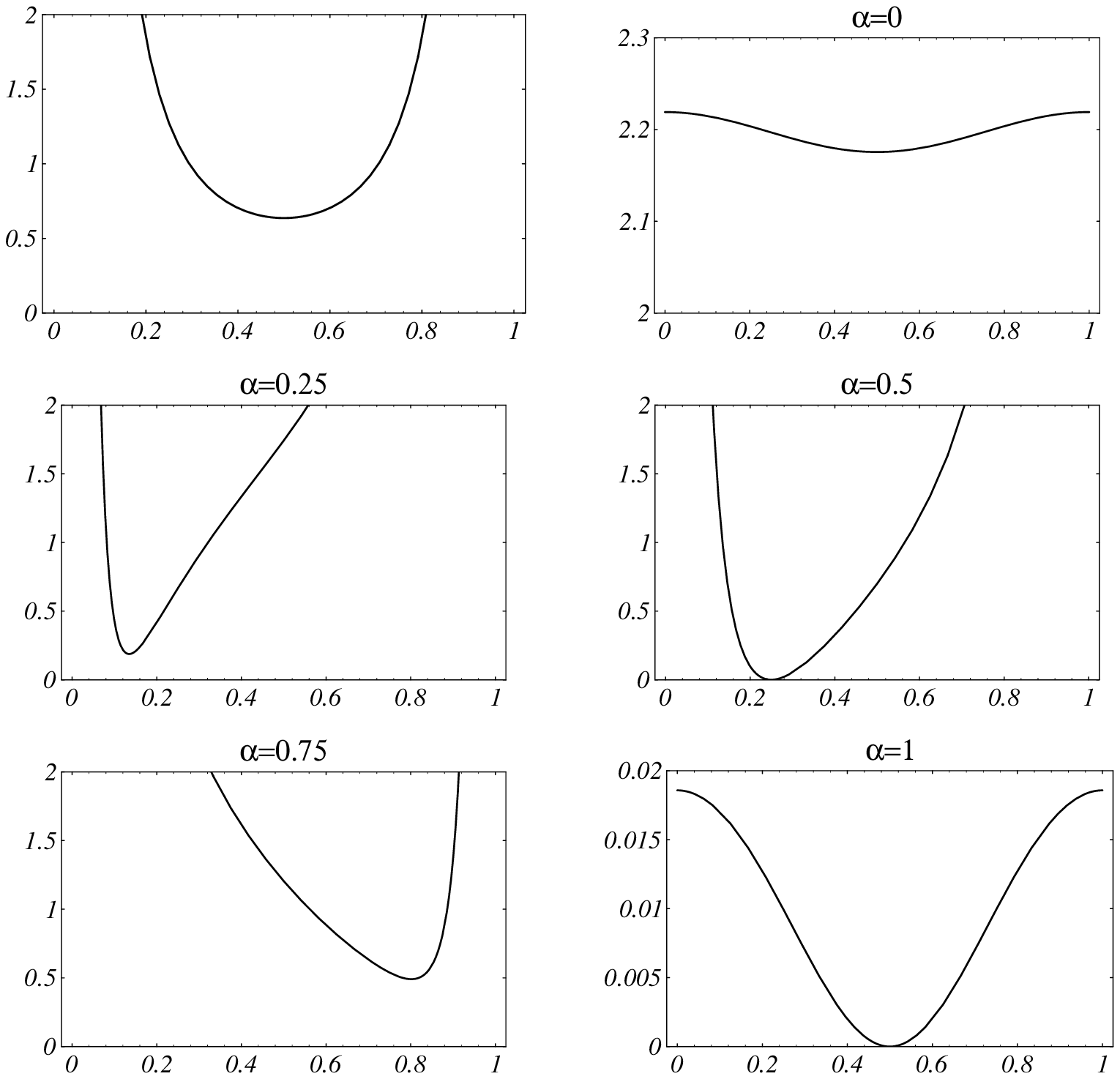}{square well cross--sections}{Displayed are the 
differential cross sections $\frac{d\sigma}{d\theta}$ per unit length
($y$-axis) in dependence of the scattering angle $\theta$ in units of
$\pi$ ($x$-axis). The first figure shows the symmetrized Aharonov-Bohm
cross-section divided by $\sin^2\pi\alpha$ for non-integral values of
$\alpha$. The other graphs depict the differential cross-section when
the statistical interaction as well as the square well potential are
present. The depth of the square well is chosen to be $V_0=25$ and the
radius is set to  $d=1$.} 
\newpage
\bildeinbindencm{14cm}{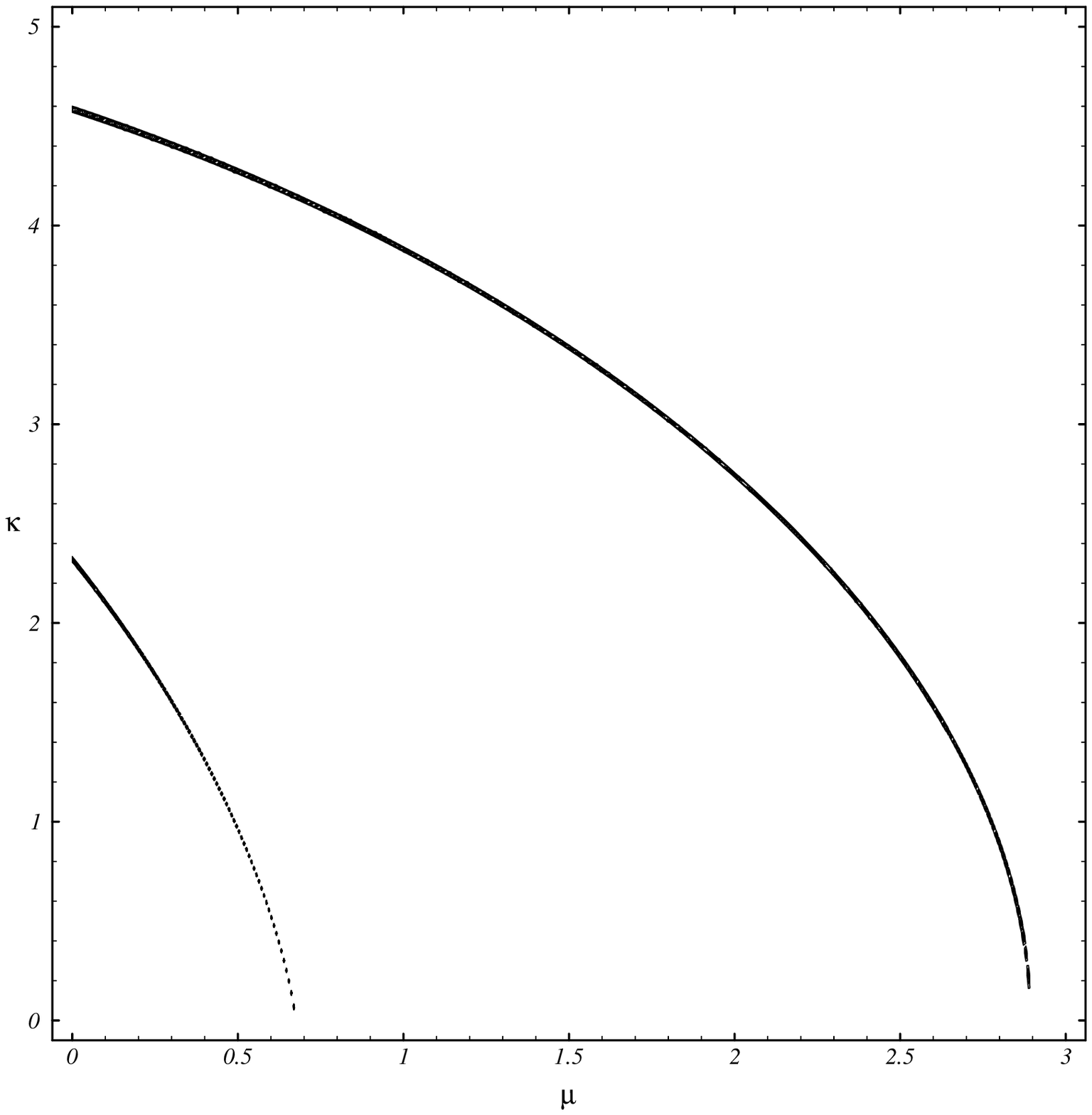}{Regge trajectories for the square well}{The graphics shows the 
zeroes of the Jost function $F$ in dependence of the momentum
$k=i\kappa,\kappa>0$ and the angular momentum $\mu$ for the square
well potential. The line broadening is due to numerical error. The
chosen numerical values are $d=1,V_0=25$. One observes two bound
states the first of which disappears at $\mu=0.674$ and the second at
$\mu=2.893$.}  

\end{document}